\documentclass[10pt, twocolumn, twoside]{IEEEtran} 
  
\usepackage{color, url, array}
\usepackage[numbers, square, comma, sort&compress]{natbib}
\usepackage{algorithm, algpseudocode}
\usepackage{graphics, epstopdf, graphicx, epsfig, psfrag, subfigure}
\usepackage{amssymb, amsmath}
\usepackage{multirow}
\usepackage{verbatim}

\newtheorem{theorem}{Theorem}[section]

\newtheorem{lemma}[theorem]{Lemma}
\newtheorem{proposition}[theorem]{Proposition}

\newtheorem{definition}{Definition}
\newtheorem{example}{Example}

\newcommand{\ie}{{\em i.e., }}
\newcommand{\eg}{{\em e.g., }}

\algnewcommand\algorithmicswitch{\textbf{switch}}
\algnewcommand\algorithmiccase{\textbf{case}}
\algdef{SE}[SWITCH]{Switch}{EndSwitch}[1]{\algorithmicswitch\ #1\ \algorithmicdo}{\algorithmicend\ \algorithmicswitch}
\algdef{SE}[CASE]{Case}{EndCase}[1]{\algorithmiccase\ #1\ \textbf{:}}{\algorithmicend\ \algorithmiccase}
\algtext*{EndSwitch}
\algtext*{EndCase}

\algnotext{EndIf}
\algnotext{EndWhile} 
\algnotext{EndFor}

\begin{document}

\title{Active Learning of Multiple Source Multiple Destination Topologies}

\author{Pegah~Sattari,~\IEEEmembership{Member,~IEEE,}
        Maciej~Kurant,
        Animashree~Anandkumar,~\IEEEmembership{Member,~IEEE,}
        Athina~Markopoulou,~\IEEEmembership{Senior~Member,~IEEE,}
        and~Michael~Rabbat,~\IEEEmembership{Member,~IEEE}
\thanks{Copyright (c) 2014 IEEE. Personal use of this material is permitted. However, permission to use this material for any other purposes must be obtained from the IEEE by sending a request to pubs-permissions@ieee.org.}
\thanks{P. Sattari was with the EECS Department, UC Irvine, CA 92697, USA, when this work was conducted. She is now with Jeda Networks Inc., Newport Beach, CA 92660, USA. E-mail: psattari@alumni.uci.edu.}
\thanks{M. Kurant was with Calit2, University of California, Irvine, CA 92697, USA, when this work was conducted. He is now with Google, Zurich, Switzerland. E-mail: maciej.kurant@gmail.com.}
\thanks{A. Anandkumar and A. Markopoulou are with the EECS Department, UC Irvine, CA 92697, USA. E-mail: a.anandkumar@uci.edu; athina@uci.edu.}
\thanks{M. Rabbat is with the Department of Electrical and Computer Engineering, McGill University, Montreal, QC, Canada. E-mail: michael.rabbat@mcgill.ca.}
\thanks{This work has been supported by NSF Award 1028394, AFOSR Award FA9550-10-1-0310 and AFOSR MURI FA9550-09-0643. The work of M. Rabbat was funded in part by the Natural Sciences and Engineering Research Council of Canada.}}

\maketitle

\begin{abstract}
\boldmath
We consider the problem of inferring the topology of a network with $M$ sources and $N$ receivers (hereafter referred to as an $M$-by-$N$ network), by sending probes between the sources and receivers. Prior work has shown that this problem can be decomposed into two parts: first, infer smaller subnetwork components (\ie $1$-by-$N$'s or $2$-by-$2$'s) and then merge these components to identify the $M$-by-$N$ topology. In this paper, we focus on the second part, which had previously received less attention in the literature. In particular, we assume that a $1$-by-$N$ topology is given and that all $2$-by-$2$ components can be queried and learned using end-to-end probes. The problem is {\em which} $2$-by-$2$'s to query and {\em how} to merge them with the given $1$-by-$N$, so as to exactly identify the $2$-by-$N$ topology, and optimize a number of performance metrics, including the number of queries (which directly translates into measurement bandwidth), time complexity, and memory usage. We provide a lower bound, $\lceil \frac{N}{2} \rceil$, on the number of $2$-by-$2$'s required by any active learning algorithm and propose two greedy algorithms. The first algorithm follows the framework of multiple hypothesis testing, in particular Generalized Binary Search (GBS), since our problem is one of active learning, from $2$-by-$2$ queries. The second algorithm is called the Receiver Elimination Algorithm (REA) and follows a bottom-up approach: at every step, it selects two receivers, queries the corresponding $2$-by-$2$, and merges it with the given $1$-by-$N$; it requires exactly $N-1$ steps, which is much less than all $\binom{N}{2}$ possible $2$-by-$2$'s. Simulation results over synthetic and realistic topologies demonstrate that both algorithms correctly identify the $2$-by-$N$ topology and are near-optimal, but REA is more efficient in practice.
\end{abstract}


\section{Introduction}

\IEEEPARstart{K}{nowledge} of network topology is important for network management, diagnosis, operation, security, and performance optimization \cite{journal, probing, merging, survey, anima-phylo, infocom2012}. In this paper, we consider a tomographic approach to topology inference, which assumes no cooperation from intermediate nodes and relies on end-to-end probes to infer internal network characteristics, including topology \cite{survey}. Typically, multicast or unicast probes are sent and received between sets of sources and receivers at the edge of the network, and the topology is inferred based on the number and order of received probes, or more generally, using some metric or correlation structure. An important performance metric is measurement bandwidth overhead: it is desirable to accurately infer  the topology using a small number of probes.
 
In this paper, we focus on the problem of multiple-source multiple-destination  topology inference: our goal is to infer the internal network ($M$-by-$N$) topology by sending probes between $M$ sources and $N$ receivers at the edge of the network. Prior work \cite{merging, probing, journal} has shown that  this problem can be decomposed into two parts: first, infer smaller subnetwork components (\eg multiple $1$-by-$N$'s or $2$-by-$2$'s) and then merge them to identify the entire $M$-by-$N$ topology. 

Significant progress has been made over the past years on the decomposition as well as the first part of the problem, \ie inferring smaller components ($1$-by-$N$'s or $2$-by-$2$'s) using active probes. One body of work developed techniques for inferring $1$-by-$N$ (\ie single-source tree) topologies using end-to-end measurements \cite{sylvia, adaptive, duffield2002, castroUnicast, topologyDelay, eriksson2010, JianNi-ToN, JianNi-arxiv, anima}. Follow-up work \cite{merging,probing,journal} showed that an $M$-by-$N$ topology can be decomposed into and reconstructed from a number of two-source, two-receiver ($2$-by-$2$) subnetwork components or ``quartets''. In \cite{journal, probing}, a practical scheme was proposed to distinguish between some quartet topologies using back-to-back unicast probes. In our recent work \cite{netcod, topologyJournal}, we proposed a method to exactly identify the topology of a quartet in networks with multicast and network coding capabilities. 

In this paper, we focus on the second part of the problem, namely selecting and merging smaller subnetwork components to exactly identify the $M$-by-$N$, which has received significantly less attention than the first part. Existing  approaches developed for merging the quartets \cite{merging, journal} have several limitations, including not  being able to exactly identify the $M$-by-$N$ topology and/or being inefficient (\eg requiring to send probes over all $\binom{N}{2}$  possible quartets). In this paper,  we formulate the problem as active learning, characterize its complexity, and follow principled approaches to design efficient algorithms to solve it. This complexity is important from both theoretical (a fundamental property of the topology inference problem) and  practical (it determines the measurement bandwidth overhead, running time and memory usage) points of view. These costs can become particularly important when we need to infer large or dynamic topologies using active measurements, and an efficient algorithm is required for that.\footnote{Examples of networks where up-to-date topology information, and thus dynamic mapping of the topology, is required include the following: detection systems that detect Internet faults \cite{failureDetect1, failureDetect2} or prefix hijacks \cite{prefixHijack} and require frequent measurements of Internet paths; content distribution networks that need to continuously monitor the topology in order to select the {\em best} content server for user requests \cite{contentDistribution}; and overlay networks that need to monitor the topology to select the best overlay routing \cite{overlayNetworks}.}

More specifically, we start from the problem of $2$-by-$N$ topology inference, which is an important special case and can then be used as a building block for inferring an $M$-by-$N$. Consistently with \cite{journal}, we assume that a (static) $1$-by-$N$ topology is known (\eg using one of the methods in \cite{sylvia, adaptive, duffield2002, castroUnicast, topologyDelay, eriksson2010, JianNi-ToN, JianNi-arxiv, anima, survey, allerton}). Then we query the quartet component by sending end-to-end probes between the two sources and the two receivers, and we learn its topology using some of the methods in \cite{probing, journal, netcod, topologyJournal, jaggi, skitter, cheswick, govindan, rocketFuel}\footnote{Other techniques may also be developed in the future as this is still an active research area. However, this is out of the scope of this paper (see Section~\ref{sec-statement}).}. The problem then becomes one of active learning: ``{\em which} quartets to query and {\em how} to merge them with the given $1$-by-$N$, so as to exactly identify the $2$-by-$N$ topology and optimize a number of performance metrics, including the number of queries (thus the measurement bandwidth), time complexity, and memory usage.'' Our contributions are as follows: 

1) We provide a lower bound of $\lceil \frac{N}{2} \rceil$ on the number of quartets required by {\em any} active learning algorithm in order to identify the $2$-by-$N$. This characterizes the inherent complexity of the problem and also serves as a rough baseline for assessing the performance of practical algorithms.

2) We formulate the problem within the framework of multiple hypothesis testing and develop an active learning algorithm based on Generalized Binary Search (GBS). This is the natural framework to pose the problem; however, we evaluate the performance of this algorithm via simulation and show that the computational complexity is high in practice.

3) As an alternative, we design an efficient Receiver Elimination Algorithm (REA), which follows a greedy bottom-up approach and provably identifies the $2$-by-$N$ topology by querying exactly $N-1$ quartets. From the active probing perspective, this is attractive since only $N-1$ queries are required, which is much lower than all $\binom{N}{2}$ possible quartets one could query. This directly results in very low measurement bandwidth, which is the main performance metric in active monitoring.

We compare the two algorithms to each other and to the lower bound via simulation over both synthetic and realistic topologies. The results show that both algorithms can exactly identify the topology and are near-optimal in terms of active measurement bandwidth. Between the two, the Receiver Elimination Algorithm is found to be very efficient in terms of running time and memory usage, and is, therefore, recommended for practical implementation.

The rest of the paper is organized as follows. Section~\ref{sec-related} summarizes related work. Section~\ref{sec-statement} provides the problem statement and terminology. Section~\ref{sec-lowerBound} provides a lower bound on the number of quartets required by any algorithm. Section~\ref{sec-gbs} proposes a greedy algorithm based on the GBS framework and evaluates its performance via simulation. Section~\ref{sec-algorithm} proposes the greedy REA, analyzes its correctness and performance, and compares it to GBS in simulation. Section~\ref{sec-discussion} discusses possible extensions. Section~\ref{sec-conclusion} concludes the paper.

\section{Related Work}
\label{sec-related}

There is a large body of prior work on inference of network topology. The most closely related to this paper are the ones using active measurements and network tomography. 

{\bf Tomographic approaches.} A survey of {\em network tomography} can be found in \cite{survey}. Tomographic approaches rely only on end-to-end measurements to infer internal network characteristics, which may include link-level parameters (such as loss and delay metrics) or the network topology \cite{shih2007, tsang2003, shih2003, deng2012, liang2003, chen2010}. In this paper, we focus on inferring the network topology. Most tomographic approaches rely on probes sent from a single source in a tree topology \cite{sylvia, adaptive, duffield2002, castroUnicast, topologyDelay, eriksson2010, JianNi-ToN, JianNi-arxiv, anima} and feed the number, order, or a monotonic property of received probes as input to statistical signal-processing techniques. 

In \cite{journal, probing, merging}, the authors formulated the multiple source multiple destination ($M$-by-$N$) tomography problem by sending probes between $M$ sources and $N$ receivers. It was shown that an $M$-by-$N$ network can be decomposed into a collection of $2$-by-$2$ components, also referred to as quartets \cite{anima-phylo, infocom2012}. Coordinated transmission of back-to-back unicast probes from the two sources and packet arrival order measurements at the two receivers were used to infer some information about the quartet topology. Assuming knowledge of $M$ $1$-by-$N$ topologies and the quartets, it was also shown how to merge a second source's $1$-by-$N$ tree topology with the first one. The resulting $M$-by-$N$ topology is not exact, but bounds were provided on the locations of the points where the two $1$-by-$N$ trees merge with each other. This approach also requires a large number of probes for statistical significance, similar to many other methods \cite{topologyDelay, adaptive, castroUnicast, sylvia, duffield2002}. Compared to \cite{journal}, our work is different in that (i) we assume perfect knowledge of the quartets, thus we identify the topology accurately; (ii) we focus on the efficiency of active learning, \ie selecting and merging the quartets, which has not been studied before.
To the best of our knowledge, the only other merging algorithm proposed in the literature is \cite{merging, journal}. However, the merging was not efficient since all possible quartets were queried exhaustively. 

In our prior work \cite{netcod, topologyJournal}, we revisited the problem of topology inference using end-to-end probes in networks where intermediate nodes are equipped with multicast and network coding capabilities. We built on \cite{journal} and extended it, using network coding at intermediate nodes to deterministically distinguish among all possible quartet topologies, which was not possible before. While in \cite{netcod, topologyJournal}, we focused on inferring the quartets fast and accurately, here we assume that any quartet can be queried and learned, and focus on efficiently selecting and merging the quartets to infer the larger topology. To the best of our knowledge, this work is the first to look at this aspect of the problem.

There also exists a rich body of work on {\em multiple hypothesis testing} for active learning problems where queries are selected adaptively. One of the contributions of this paper is to formulate this problem in that framework and design an algorithm based on one such active learning scheme, GBS \cite{gbs-nowak, dasgupta, castroNowak}, which we describe in detail in Section~\ref{sec-gbs}. 

Topology inference problems have also been studied in the context of {\em phylogenetic} trees \cite{phylogenetic, pearl}. The work in \cite{infocom2012} built on \cite{pearl} and proposed robust algorithms for multiple source tree topology inference. The work in \cite{anima-phylo} inferred the topology of sparse random graphs using end-to-end measurements between a small subset of nodes. However, the quartet structures and the way we measure them are different in our case due to the nature of active probing in network tomography (see problem formulation in Section~\ref{sec-statement}).

{\bf Traceroute-based approaches.} An alternative to tomographic approaches is {\tt traceroute}-based techniques, which rely on cooperation of nodes in the middle of the network, in order to connect the ids of nodes along paths and reconstruct the topology \cite{skitter, cheswick, govindan, rocketFuel}. These approaches face their own set of challenges: not all intermediate nodes cooperate by responding, many of them have multiple network interfaces (ids), and {\tt traceroute} is often turned off for security reasons. Therefore, {\tt traceroute}-based methods must deal with missing or incomplete data and alias problems. Regardless, the point of this paper is not to compare the tomographic approaches against the {\tt traceroute}-based approaches, but to provide active learning algorithms that can probe/query the network in an efficient way. Querying the network can be achieved via end-to-end probes, {\tt traceroute}, or even in a passive way. As long as we can query $2$-by-$2$ components, the active learning approach should be applicable and useful in minimizing the cost of all such approaches.

{\bf Relation to the conference version.} This journal paper builds on our conference paper in \cite{ciss}. In addition to revisions and elaborating on parts of the writing, new materials/contributions in this paper include the following: the formulation of the problem in the GBS framework as well as the performance evaluation of the GBS algorithm via simulation, and its comparison against REA.

\section{Problem Statement}
\label{sec-statement}

{\bf $\boldsymbol M$-by-$\boldsymbol N$ Topology to be inferred.} Consider an $M$-by-$N$ topology as a directed acyclic graph (DAG), between $M$ source nodes $\mathcal{S}=\{S_1,...,S_M\}$ and $N$ receiver nodes $\mathcal{R}=\{R_1,...,R_N\}$. We denote this $M$-by-$N$ topology by $G_{\mathcal{S}\times\mathcal{R}}$. Note that $G_{S_i\times\mathcal{R}}$, $i=1,...,M$, is a $1$-by-$N$ tree. Similar to \cite{journal, probing, merging}, we assume that a predetermined routing policy maps each source-destination pair to a unique route from the source to the destination. This implies the following three properties, first stated in \cite{journal}:\footnote{These assumptions are realistic, the same as in \cite{probing, merging, journal}, and consistent with the destination-based routing used in the Internet: each router decides the next hop taken by a packet using a routing table lookup on the destination address. We further assume that the network does not employ load balancing.}
\begin{itemize}
\item[A1] For every source $S_i$ and every receiver $R_j$, there is a unique path $P_{ij}$.
\item[A2] Two paths $P_{ij}$ and $P_{ik}$, $j\neq k$, branch at a {\em branching point} $B$, and they never merge again.
\item[A3] Two paths $P_{ik}$ and $P_{jk}$, $i\neq j$, merge at a {\em joining point} $J$, and they never split again.
\end{itemize}

\begin{figure}[t!]
\centering
\includegraphics[scale=0.29]{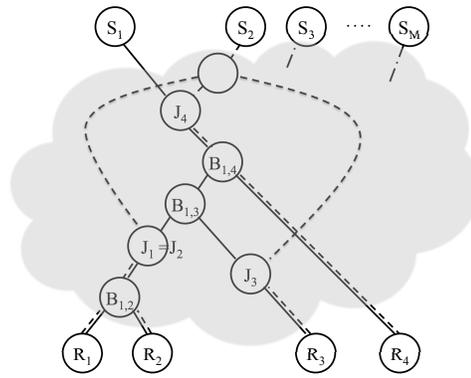}
\caption{An example $2$-by-$4$ topology. The solid lines and branching points $B_{i,j}$'s depict the $S_1$ tree topology, $G_{S_1\times\mathcal{R}}$. $J_i$ is a joining point, where $P_{2i}$ (indicated by the dashed lines) joins $G_{S_1\times\mathcal{R}}$. An example quartet is the part of the network connecting $S_1,S_2$ to $R_1,R_2$, which is type 1 since both $J_1$ and $J_2$ lie above the branching point of $R_1$ and $R_2$ in $G_{S_1\times\mathcal{R}}$, \ie $B_{1,2}$.} 
\label{fig-problemStatement}
\end{figure}

\begin{figure*}[t!]
\centering
\subfigure[type 1]{\includegraphics[scale=0.18]{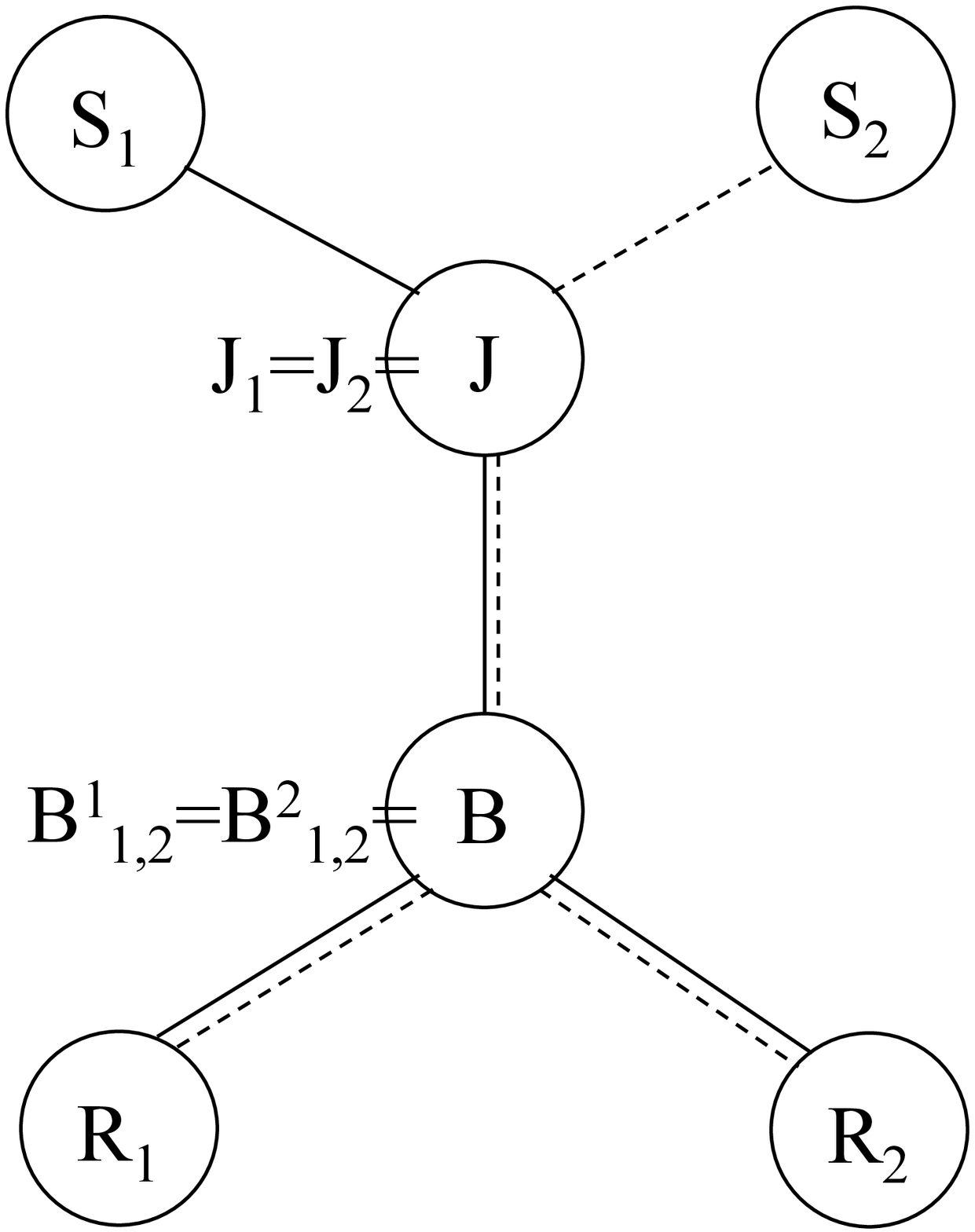}}
\hspace{1cm}
\subfigure[type 2]{\includegraphics[scale=0.18]{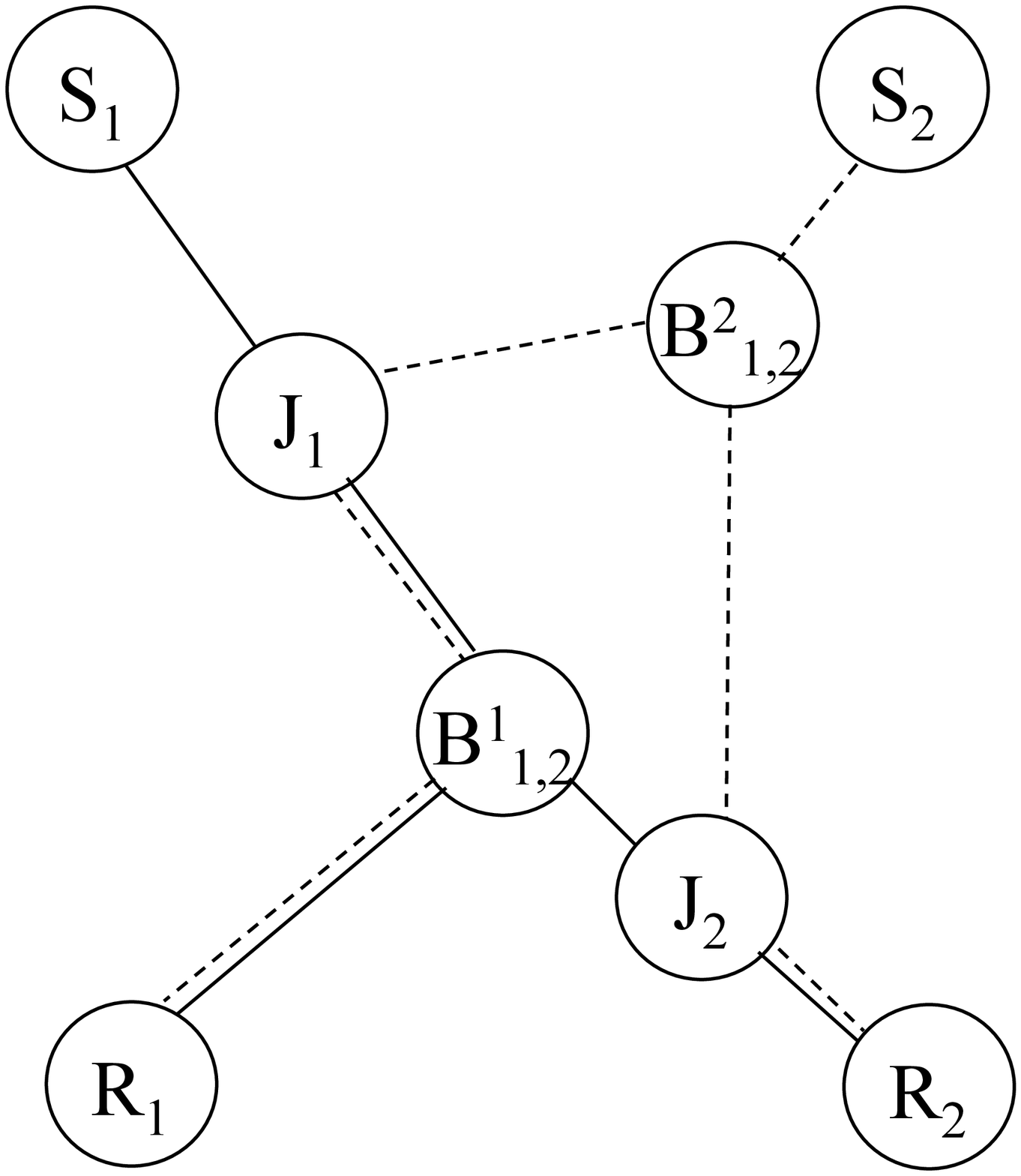}}
\hspace{1cm}
\subfigure[type 3]{\includegraphics[scale=0.18]{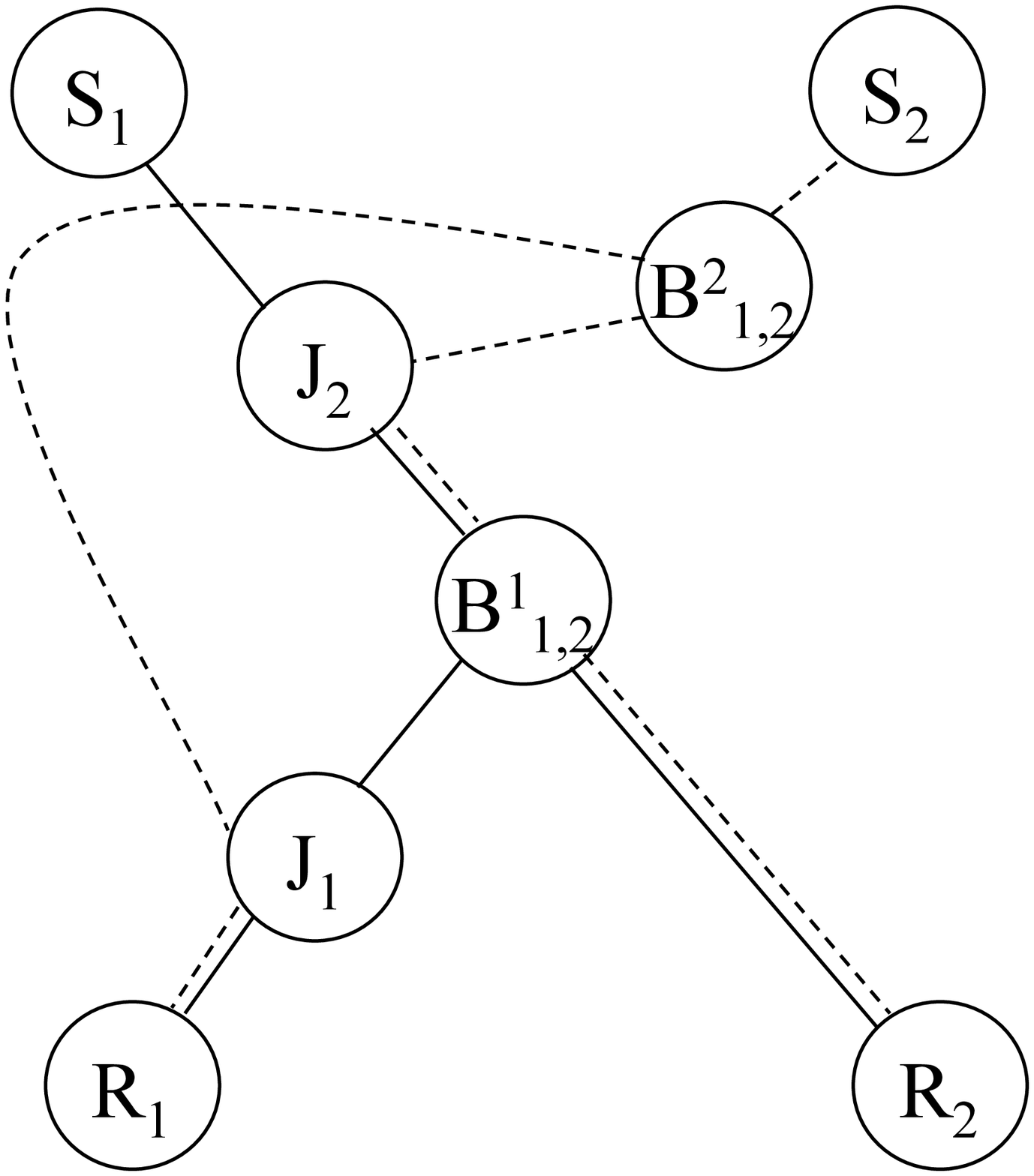}}
\hspace{1cm}
\subfigure[type 4]{\includegraphics[scale=0.18]{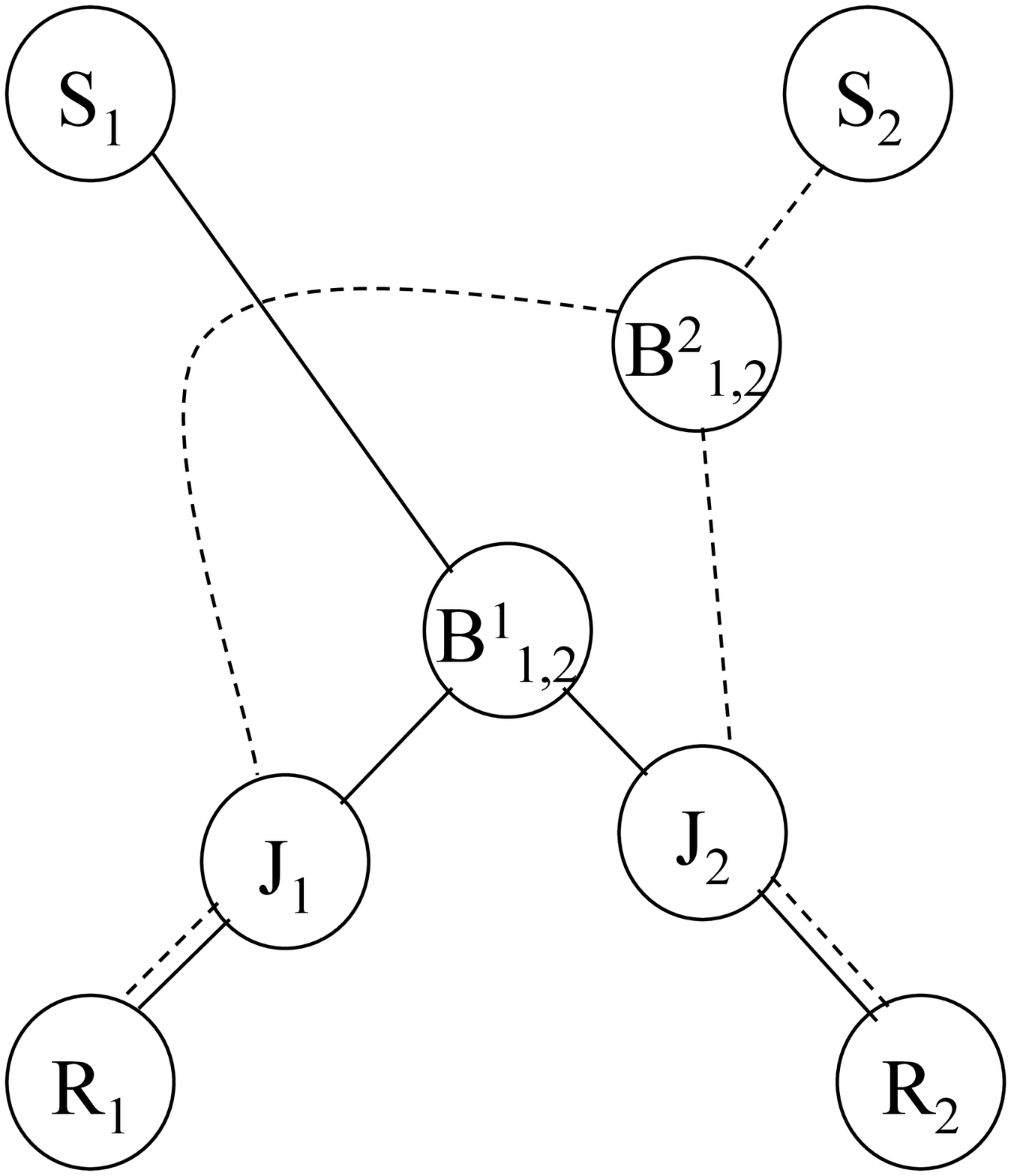}}
\caption{The four possible types of a quartet ($2$-by-$2$ subnetwork component). There are two sources $S_1$ and $S_2$ multicasting packets $x_1$ and $x_2$, respectively, to two receivers $R_1$ and $R_2$. All links are directed downwards, but arrowheads are omitted to avoid cluttering. The $1$-by-$2$ topology of $S_1$ is a tree composed of $S_1, B^1_{1,2}, R_1, R_2$. Similarly, the $1$-by-$2$ tree rooted at $S_2$ is $S_2, B^2_{1,2}, R_1, R_2$. $J_1$ and $J_2$ are joining points, where the paths from $S_2$ to $R_1$ and $R_2$ join/merge with $S_1$'s tree topology.} 
\label{fig-2by2} 
\end{figure*}

We are interested in inferring the logical topology\footnote{A logical topology is obtained from a physical topology by ignoring nodes with in-degree = out-degree = 1. Such nodes cannot be identified and network tomography always focuses on inferring logical topologies.}, defined by the branching and joining points defined above. We present most of our discussion in terms of $M = 2$, \ie inferring a $2$-by-$N$ topology $G_{\mathcal{S}\times\mathcal{R}}$, $\mathcal{S}=\{S_1,S_2\}$; an $M$-by-$N$ topology, $\mathcal{S}=\{S_1,...,S_M\}$, can then be constructed by merging smaller structures, as we describe in Section~\ref{sec-discussion}. 

\begin{example} Fig.~\ref{fig-problemStatement} illustrates an example $2$-by-$N$ topology with $N=4$. The logical tree topology of $S_1$ is shown by solid lines and branching points $B_{i,j}$'s. Each $J_i$ depicts a joining point, where the path from $S_2$ to receiver $R_i$ (indicated by the dashed lines) joins the $S_1$ tree. For example, the path from $S_2$ to $R_1$ joins the $S_1$ tree at a point between $B_{1,3}$ and $B_{1,2}$, whereas the path to $R_4$ joins at a point above $B_{1,4}$. 
\hfill{$\blacksquare$}
\end{example}

{\bf Quartet Components.} In \cite{journal}, it has been shown that an $M$-by-$N$ topology can be decomposed into a collection of $2$-by-$2$ subnetwork components, which, in this paper, we call {\em quartets}, following the terminology in \cite{anima-phylo, infocom2012}. Each quartet can be of four possible types, as shown in Fig.~\ref{fig-2by2}. We refer to Fig.~\ref{fig-2by2} (a), (b), (c), and (d) as types 1, 2, 3, and 4, respectively. Note that in type 1, the joining points for both receivers coincide ($J_1\equiv J_2$) and the branching points for both sources coincide ($B^1_{1,2}\equiv B^2_{1,2}$). However, the other three types (2, 3, and 4), have two distinct joining points and two distinct branching points. 

In order to infer the type of a quartet between two sources $S_1, S_2$ and two receivers $R_i, R_j$, a set of probes must be sent from $S_1, S_2$ to $R_i, R_j$. The received probes can then be processed using techniques such as the ones developed in: \cite{journal, probing} (which distinguish type 1 from types 2, 3, 4 by sending back-to-back unicast probes); \cite{netcod, topologyJournal} (which distinguish among all four types exploiting multicast and network coding); \cite{jaggi} (which can exactly infer the topology of a super-source to two receivers using network coding); {\tt traceroute}  \cite{cheswick, skitter, govindan, rocketFuel} from the two sources to the two receivers; or other techniques that may be developed in the future, since this is still an active research area. We consider the design of these techniques to be out of the scope of this paper and we focus on their use by active learning algorithms to perform a {\em query}, \ie to learn a quartet type by sending and processing a set of active probes.

Being able to query the type of a quartet enables inference of an $M$-by-$N$ topology in two steps, as follows: first infer the type of each quartet, and then merge these quartets to identify the original topology. Indeed, knowing the type of the quartet, we can use Fig.~\ref{fig-2by2} to infer the relative location of joining and branching points. For example, knowing that the quartet is of type 1 implies that (i) the two joining points coincide $J_1\equiv J_2$, (ii) the two branching points coincide $B^1_{1,2}\equiv B^2_{1,2}$, and (iii) the joining point is above the branching point. Similar inferences can be made from the other types.

{\bf Problem Statement.}
Consistently with \cite{journal}, we assume that $G_{S_1\times\mathcal{R}}$ (\ie the $1$-by-$N$ tree topology rooted at $S_1$, which contains only branching points) is known (\eg using one of the methods in \cite{sylvia, adaptive, duffield2002, castroUnicast, topologyDelay, eriksson2010, JianNi-ToN, JianNi-arxiv, anima, survey, allerton}). We also assume that the type of the quartet between $S_1$, a new source $S_2$, and any two receivers can be queried and learned, as explained above.

Given (i) $G_{S_1\times\mathcal{R}}$ and (ii) the ability to query the quartet type between $S_1$, $S_2$, and any two receivers $R_i$, $R_j$, our goal is to {\em identify} all joining points, $\mathcal{J}_N=\{J_1,J_2,...,J_N\}$, where the paths from $S_2$ to each receiver join the tree describing paths from $S_1$ to the same set of receivers.\footnote{Note that we do not need to identify the branching points of $S_2$ because the tree topology of $S_2$, like $S_1$, is given. We are only interested in identifying where these two tree topologies join/merge with each other, \ie we only want to identify the joining points of $S_1$ and $S_2$ trees.} Identifying a  joining point $J_i$ (for receiver $R_i$) means locating $J_i$ on a single logical link, between two branching points on $G_{S_1\times\mathcal{R}}$. {\em E.g.,} in Fig.~\ref{fig-problemStatement}, the path from $S_2$ to $R_1$ joins the $S_1$ tree at a point between nodes $B_{1,3}$ and $B_{1,2}$; \ie $J_1$ is located on the link $(B_{1,3},B_{1,2})$. 

We achieve this goal via active learning:  we start from the given, static, $1$-by-$N$ topology $G_{S_1\times\mathcal{R}}$, and proceed by updating it in steps. In each step, we select which quartet to query (\ie which two receivers to send probes to, from sources $S_1$ and $S_2$)\footnote{Since we focus on $M=2$, \ie only two sources $S_1$ and $S_2$, we represent the quartets $(S_1,S_2,R_i,R_j)$ only by the receivers $(R_i,R_j)$ for brevity.}, and learn its type (after sending and processing the received probes, we have essentially queried and learned the type of that quartet). We then merge this quartet with the known topology so far.  We continue until identifying the entire $2$-by-$N$. The goal is to exactly identify the $2$-by-$N$ topology while minimizing the number of queries (\ie set of probes sent to measure the quartets). This metric is important because it directly translates into measurement bandwidth. Additional performance metrics that it is desirable to keep low include:  merging complexity and memory usage.

\section{Lower Bound}
\label{sec-lowerBound}

First, we provide a lower bound on the number of quartets required by any active learning algorithm to identify the $2$-by-$N$ topology. This lower bound clearly depends on the topology we want to identify and serves as a baseline for assessing the performance of the proposed algorithms.
\begin{theorem}
\label{theorem-minimum2by2s}
Given $G_{S_1\times\mathcal{R}}$, the number of quartets required to be queried by any  algorithm in order to identify all the joining points in $G_{\mathcal{S}\times\mathcal{R}}$, $\mathcal{S}=\{S_1,S_2\}$, is at least $\lceil \frac{N}{2} \rceil$.
\end{theorem}

Before proving the theorem, let us discuss some examples that illustrate the intuition and that this bound is not  tight.  

\begin{figure}[t!]
\centering 
\subfigure[Two quartets are sufficient.]{\includegraphics[scale=0.21]{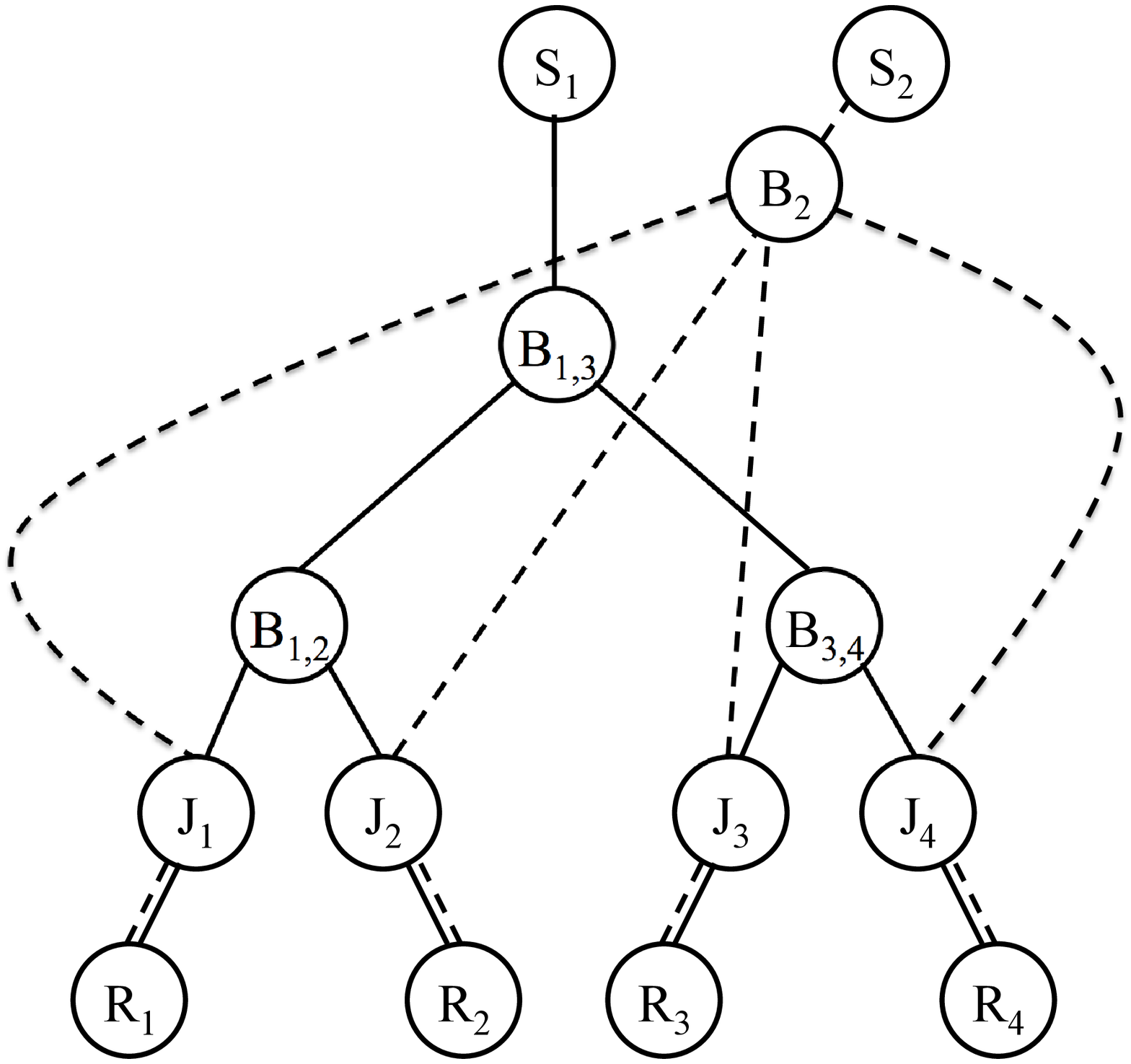}} 
\hspace{0cm}  
\subfigure[Three quartets are required.]{\includegraphics[scale=0.21]{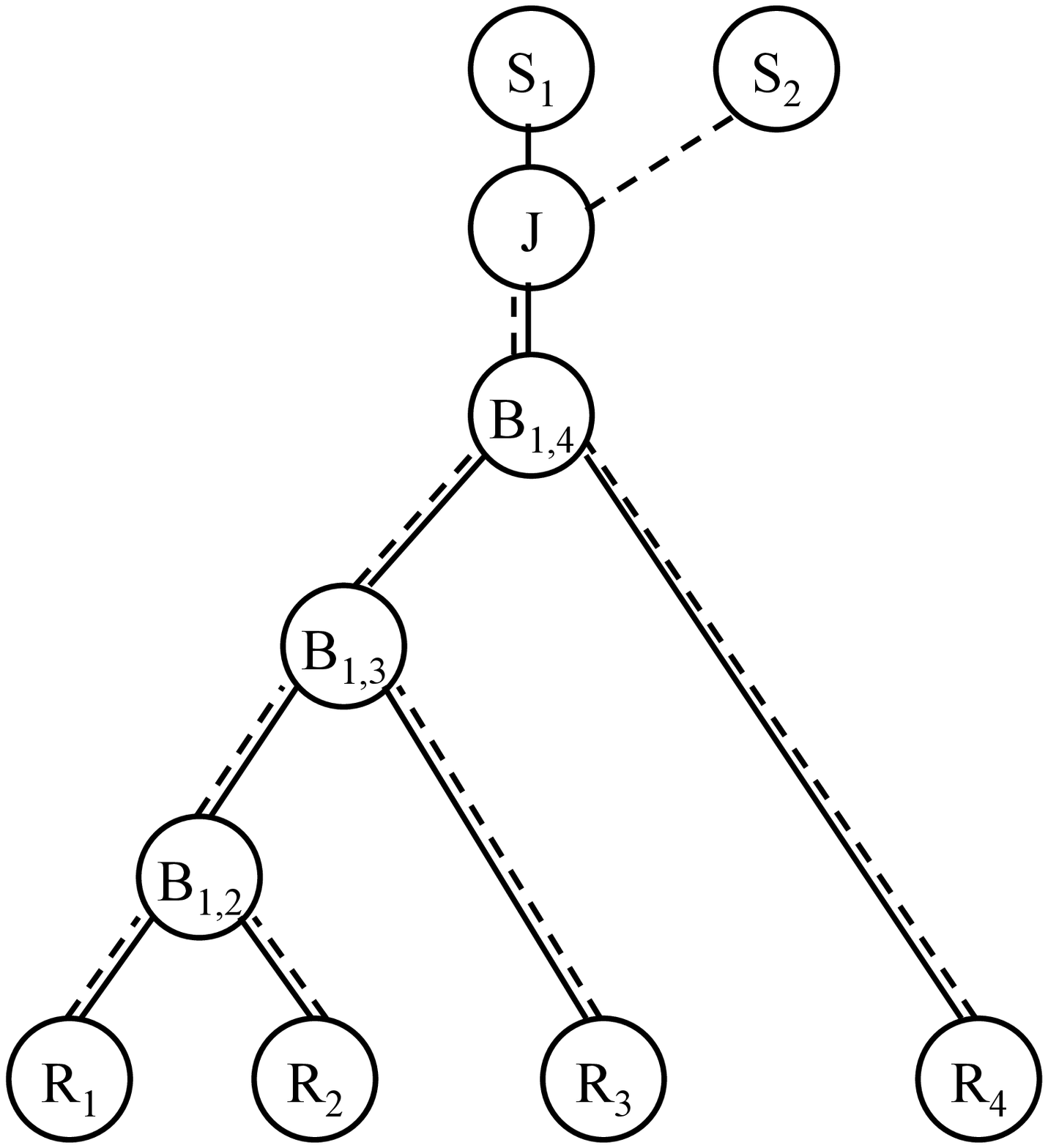}} 
\caption{Two example $2$-by-$N$ topologies with $N=4$. In (a), $\frac{N}{2}$ quartets are sufficient to identify the joining points, \ie $(R_1,R_2)$ and $(R_3,R_4)$. In (b), more than $\frac{N}{2}$ quartets are required, \eg $(R_1,R_2)$, $(R_1,R_3)$, and $(R_1,R_4)$.} 
\label{fig-minimumExamples}
\end{figure}

\begin{example}
Fig.~\ref{fig-minimumExamples}(a) shows a $2$-by-$N$ topology with $N=4$, which requires querying exactly $\frac{N}{2}=2$ quartets in order to uniquely identify all the joining points. This is because, in this particular topology, knowing the types of  $(R_1,R_2)$ and $(R_3,R_4)$ is sufficient for identifying all four joining points. Indeed, $(R_1,R_2)$ is of type 4, which, according to Fig. \ref{fig-2by2}, means that both $J_1$ and $J_2$ lie below $B_{1,2}$; also $(R_3,R_4)$ is type 4, which means that both $J_3$ and $J_4$ are below $B_{3,4}$. Thus, each joining point is identified on a single logical link.
\hfill{$\blacksquare$}
\end{example}

\begin{example}
Fig.~\ref{fig-minimumExamples}(b) shows an example where $\frac{N}{2}=2$ quartets are not sufficient and 3 quartets are required to identify all the joining points. There exist $\binom{4}{2}=6$ possible quartets in this topology, from which $\binom{6}{2}=15$ pairs of quartets can be selected; one can check that none of the 15 possible pairs can uniquely identify all the joining points. For example, let us consider $(R_1,R_2)$. Since it is of type 1, Fig. \ref{fig-2by2} indicates that $J_1\equiv J_2$ and both of them lie above $B_{1,2}$. However, there is more than a single link above $B_{1,2}$; therefore, we continue by considering  $(R_1,R_3)$. It is again of type 1, which means that  $J_1\equiv J_3$ is located above $B_{1,3}$. Therefore, we go one step further and consider $(R_1,R_4)$. Since this is also of type 1, $J_1\equiv J_4$ lies above $B_{1,4}$. At this step, we only have a single link between $S_1$ and $B_{1,4}$ and thus, $J_1\equiv J_2\equiv J_3\equiv J_4$ are all identified (depicted as $J$ in Fig.~\ref{fig-minimumExamples}(b)). Although there are other choices of triplets of quartets, in this topology, at least 3 quartets are required. 
\hfill{$\blacksquare$}
\end{example}

From these examples, one can see that the lower bound of $\lceil \frac{N}{2} \rceil$ is not tight and it is not achievable in every topology. Theorem~\ref{theorem-minimum2by2s} follows from the following lemma.  

\begin{lemma}
\label{lem-minimum2by2s}
In order for an algorithm to identify all joining points for all the receivers, each receiver needs to appear in the set of quartets queried by the algorithm at least once. 
\end{lemma}
\begin{IEEEproof}
Assume that there exists a receiver $R_i$ that has not been queried in any of the quartets. We show that even with complete knowledge of all other joining points, there exist at least two possible and feasible locations for $J_i$, as follows.

{\em Location 1:} $J_i$ lies on the last incoming link to $R_i$, \ie on the link between the parent of $R_i$ in the $S_1$ tree (which from now on, we denote by $parent(R_i)$), and $R_i$.  For example in Fig.~\ref{fig-minimumExamples}(a) and Fig.~\ref{fig-minimumExamples}(b), assume that $R_i=R_2$; then Location 1 would be the link ($B_{1,2},R_2)$. This is allowed by the routing assumptions in Section~\ref{sec-statement} because (1) there is a unique path $P_{2i}$; (2) $P_{2i}$ never merges with $P_{2j}$, $j\neq i$; and (3) $P_{2i}$ merges with $P_{1i}$ at $J_i$, and they continue together until they reach $R_i$.  

{\em Location 2:} Define $J_i$ as follows. On path $P_{1i}$, start at $parent(R_i)$ and move up towards $S_1$, until the first link that does not fully overlap with any $P_{2j}$, $j\neq i$. Place $J_i$ on that link. For example in Fig.~\ref{fig-minimumExamples}(a), Location 2 for $J_2$ would be the link ($B_{1,3},B_{1,2}$); whereas in Fig.~\ref{fig-minimumExamples}(b), it would be the link ($S_1,B_{1,4})$. This location is also allowed by the assumptions in Section~\ref{sec-statement}: 
\begin{itemize} 
\item[A1] There is a unique path $P_{2i}$. 
\item[A2] For every $j\neq i$, the two paths $P_{2i}$ and $P_{2j}$ never join after they branch. Indeed, if $J_j$ is located above $J_i$ on $P_{1i}$, then this is guaranteed by the construction of $J_i$. In contrast, $J_j$ cannot be located below $J_i$ on $P_{1i}$ since this would imply the violation of A2 even before adding $J_i$.  
\item[A3] $P_{2i}$ merges with $P_{1i}$ at $J_i$ and they never split.
\end{itemize}
Thus, both Location 1 and Location 2 are valid  for $J_i$, according to the routing assumptions, and  $J_i$ cannot be uniquely identified. Therefore, $R_i$ needs to be queried at least once.
\end{IEEEproof}

Theorem~\ref{theorem-minimum2by2s} follows from the following reasoning: each quartet involves two receivers, and thus, at least $\lceil \frac{N}{2} \rceil$ quartets are required for each receiver to appear in the set of quartets queried by the algorithm at least once.

\section{A Generalized Binary Search Algorithm}
\label{sec-gbs}

\subsection{Background on GBS}
\label{sec-background}

The GBS approach has been proposed for the problem of determining a binary-valued function through a sequence of strategically selected queries, as explained in the following \cite{gbs-nowak}. Consider a finite (potentially very large) collection of binary-valued functions $\mathcal{H}$, called the ``hypothesis space'', defined on a domain $\mathcal{X}$, called the ``query space''. Each $h \in \mathcal{H}$ is a mapping from $\mathcal{X}$ to $\{+1,-1\}$. Let $|\mathcal{H}|$ denote the cardinality of $\mathcal{H}$, \ie the total number of hypotheses. The functions $h \in \mathcal{H}$ are assumed to be unique, and one function, $h^* \in \mathcal{H}$, produces the correct binary labeling. $h^*$ is assumed to be fixed but unknown. The goal is to determine $h^*$ through as few queries from $\mathcal{X}$ as possible. Therefore, the queries need to be selected strategically in a sequential manner such that $h^*$ is identified as quickly as possible.

It has been shown that the learning problem described above is NP-complete \cite{NPcomplete}; a practical heuristic has been proposed in the form of a greedy algorithm called Generalized Binary Search (GBS). At each step, GBS selects a query that results in the most even split of the hypotheses under consideration into two subsets, responding $+1$ and $-1$, respectively, to the query. The correct response to the query eliminates one of these two subsets from further consideration. The work in \cite{gbs-nowak} characterizes the worst-case number of queries required by GBS in order to identify the correct hypothesis $h^*$. The main result of \cite{gbs-nowak} indicates that under certain conditions on the query and hypothesis spaces, the query complexity of GBS (\ie the minimum number of queries required by GBS to identify $h^*$) is near-optimal, \ie within a constant factor of $\log_2 |\mathcal{H}|$. The constant depends on two parameters $c^*$ and $k$, defined in \cite{gbs-nowak}, and it is desirable that they are both as small as possible.

In this section, we pose our problem in the GBS framework and use the GBS algorithm because (i) our problem is one of active learning and lends itself naturally to be posed in the GBS framework, and (ii) GBS is a principled (although not optimal) approach with provable correctness and performance guarantees \cite{gbs-nowak}.

\subsection{Merging Logical Topologies in the GBS Framework}

In this section, we formulate our problem within the GBS framework. Consider a set of hypotheses $\mathcal{H}$, where each hypothesis $h \in \mathcal{H}$ is a configuration that results from placing each joining point $J_i$ on an arbitrary link in the path $P_{1i}$ in the $S_1$ tree. The query space $\mathcal{X}$ is the set of all queries for all the quartets, where each query $x \in \mathcal{X}$ asks about the type of a quartet $(R_i,R_j)$. Since in our problem, each such query $x$ has 4 possible answers (corresponding to the 4 quartet types), we need to modify our queries to make them consistent with the binary functions in the standard GBS framework. We assume that each query $x$ consists of 4 subqueries, each of which asks whether $(R_i,R_j)$ is of a specific type (1, 2, 3, or 4) or not; {\em i.e.}:
\begin{displaymath}
x = \left\{ \begin{array}{l}
\text{Is $(R_i,R_j)$ of type 1?}\\
\text{Is $(R_i,R_j)$ of type 2?}\\
\text{Is $(R_i,R_j)$ of type 3?}\\
\text{Is $(R_i,R_j)$ of type 4?}
\end{array} \right.
\end{displaymath}
The answer to each such subquery is binary, which is consistent with the GBS formulation. Of course, not all four subqueries are always required for a quartet; one would stop as soon as she gets the first ``yes'', which would reveal the type of the quartet. Note, however, that we count the number of queries (not subqueries) as the performance metric of the GBS algorithm.

\begin{algorithm}[t!]
\begin{footnotesize}
\caption{\label{alg-gbs} GBS algorithm for identifying the joining points.}
\begin{algorithmic}[1]
\State Let $J=[0,0,...,0]$ be a vector of length $N$, which represents the locations of the joining points.
\While{$\exists$ $0$ in $J$}
\State Let $wcB=[\,]$ represent the worst case benefits for all the quartets. 
\For{each receiver $R_i$}
\For{each receiver $R_j$, $j>i$}
\State Let $B_{i,j}$ be the lowest common ancestor of $R_i,R_j$ in $G_{S_1\times\mathcal{R}}$
\State Let $up_{i}\subset P_{1i}$ be the subset of $P_{1i}$ located above $B_{i,j}$
\State Let $up_{j}\subset P_{1j}$ be the subset of $P_{1j}$ located above $B_{i,j}$
\State Let $dn_{i}\subset P_{1i}$ be the subset of $P_{1i}$ located below $B_{i,j}$
\State Let $dn_{j}\subset P_{1j}$ be the subset of $P_{1j}$ located below $B_{i,j}$
\State \textit{type1\_B}$=\frac{|up_{i}|}{|P_{1i}||P_{1j}|}$   
\State \textit{type2\_B}$=\frac{|up_{i}||dn_{j}|}{|P_{1i}||P_{1j}|}$
\State \textit{type3\_B}$=\frac{|dn_{i}||up_{j}|}{|P_{1i}||P_{1j}|}$
\State \textit{type4\_B}$=\frac{|dn_{i}||dn_{j}|}{|P_{1i}||P_{1j}|}$
\State \textit{wcB.append(max([type1\_B, type2\_B, type3\_B, type4\_B]))}
\EndFor
\EndFor
\State \textit{selectedQuartet=wcB.index(min(wcB))}
\State Let \textit{selectedQuartetType} be the type of \textit{selectedQuartet}.
\Switch{\textit{selectedQuartetType}}
\Case{type 1}
\State $P_{1i} \longleftarrow up_{i}$
\State $P_{1j} \longleftarrow up_{j}$
\EndCase
\Case{type 2}
\State $P_{1i} \longleftarrow up_{i}$
\State $P_{1j} \longleftarrow dn_{j}$
\EndCase
\Case{type 3}
\State $P_{1i} \longleftarrow dn_{i}$
\State $P_{1j} \longleftarrow up_{j}$
\EndCase
\Case{type 4}
\State $P_{1i} \longleftarrow dn_{i}$
\State $P_{1j} \longleftarrow dn_{j}$
\EndCase
\EndSwitch
\If{$|P_{1i} |==1$}
\State $J_i=P_{1i}$
\EndIf
\If{$|P_{1j} |==1$}
\State $J_j=P_{1j}$
\EndIf
\EndWhile
\State Output $J$.
\end{algorithmic}
\end{footnotesize}
\end{algorithm}

Our goal is to find the target hypothesis $h^*$, which is the configuration that results from the correct placement of the joining points in the $S_1$ topology, using as few queries (\ie the knowledge of as few quartet types) as possible.\footnote{More formally, $h^*$ answers every query, for any pair of receivers, in accordance with the true $2$-by-$N$ topology. Mathematically, $h^*$ is a mapping from queries to $\{+1,-1\}$, not a topology itself. However, there is a bijection between all $2$-by-$N$ logical topologies and the corresponding mappings in $\mathcal{H}$, and therefore, knowing $h^*$ is equivalent to knowing the $2$-by-$N$ topology.}

Algorithm~\ref{alg-gbs} describes a greedy strategy based on GBS for determining $h^*$. In the beginning, there are $|\mathcal{H}|$ possible hypotheses. In each step, the algorithm selects the best (\ie maximally discriminating \cite{gbs-nowak}) quartet to query as follows. By querying a quartet and learning its type, some information is obtained about the locations of two joining points. Thus, the number of feasible hypotheses, which agree with the constraints imposed by the quartets queried and learned so far, is reduced by a number, which depends on the topology in general. We call this number the {\em benefit} of the quartet. The best quartet to select to query is the one with maximum benefit. However, the benefit of each quartet becomes known only \textit{after} it is queried. Thus, the algorithm considers all four possible types for every possible quartet, and focuses on the worst case benefit of that quartet, \ie the type that gives the minimum benefit. The best quartet to query is the one with maximum worst case benefit.

\begin{figure*}[t!]
\centering
\subfigure[$G_{S_1\times\mathcal{R}}$, star topology.]{\includegraphics[scale=0.20]{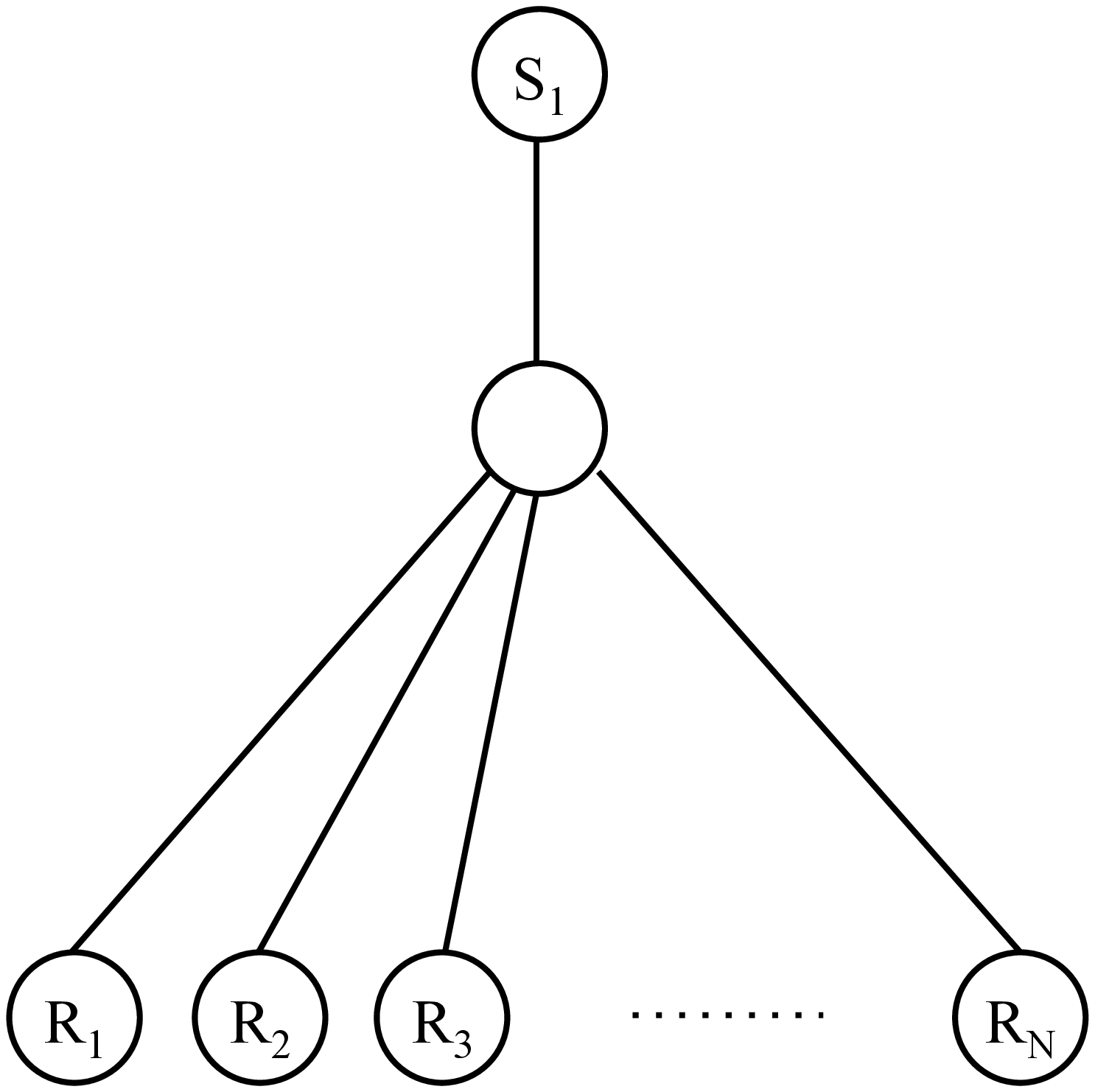}} \hspace{0cm}
\subfigure[$G_{S_1\times\mathcal{R}}$, perfect binary tree.]{\includegraphics[scale=0.20]{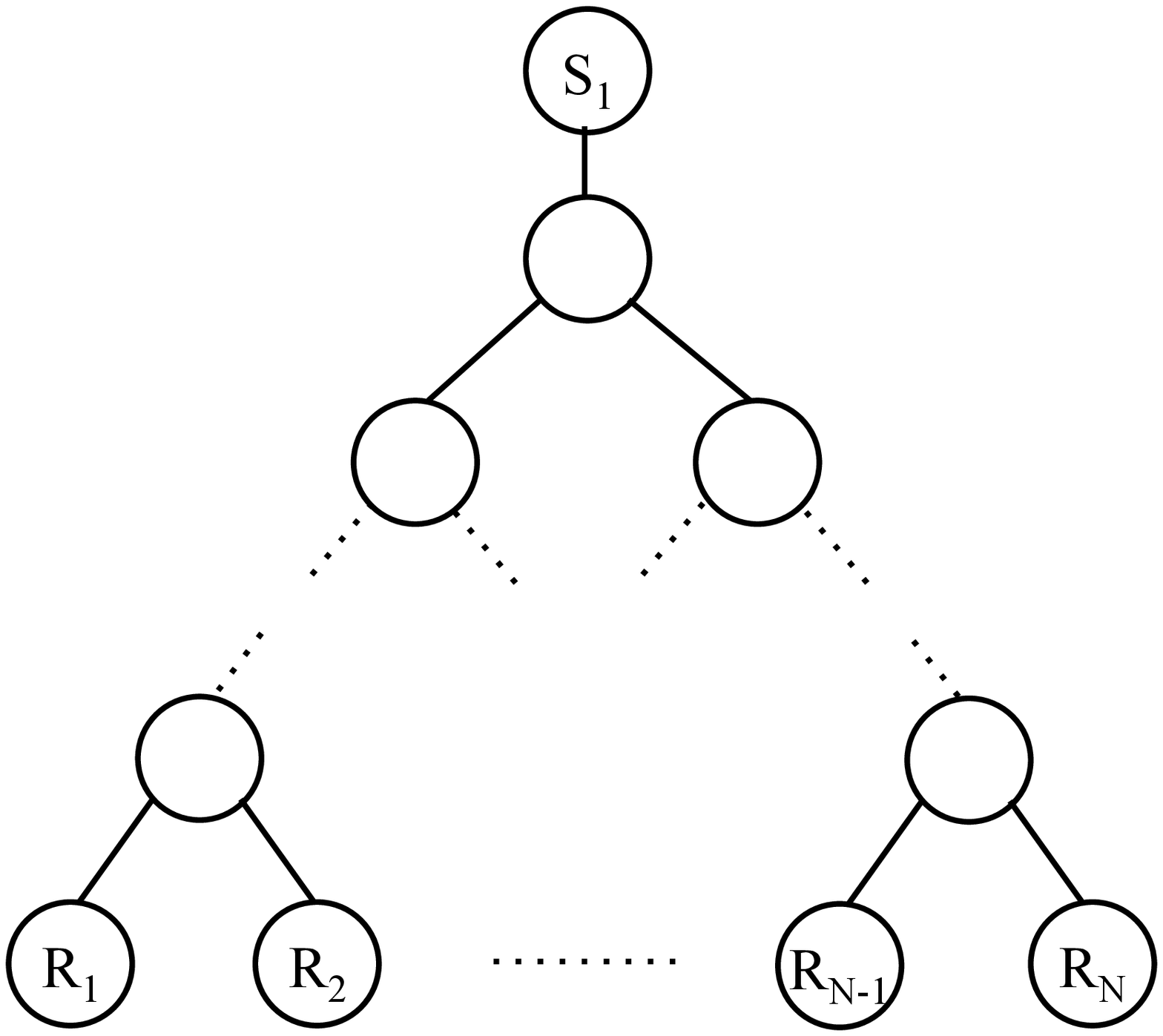}} 
\hspace{0cm}
\subfigure[$G_{S_1\times\mathcal{R}}$, tall binary tree.]{\includegraphics[scale=0.20]{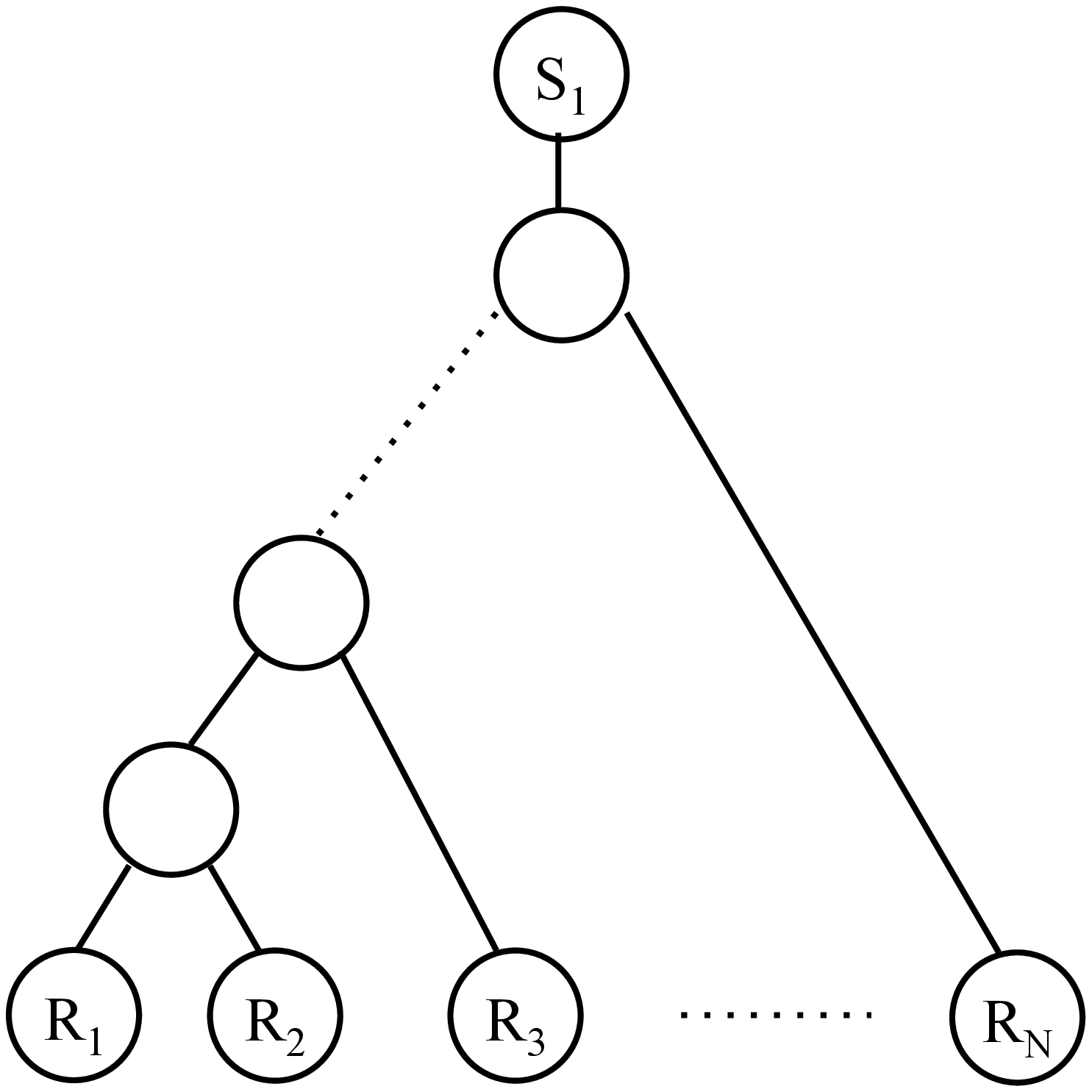}} 
\hspace{0cm}
\subfigure[$G_{S_1\times\mathcal{R}}$, perfect ternary tree.]{\includegraphics[scale=0.20]{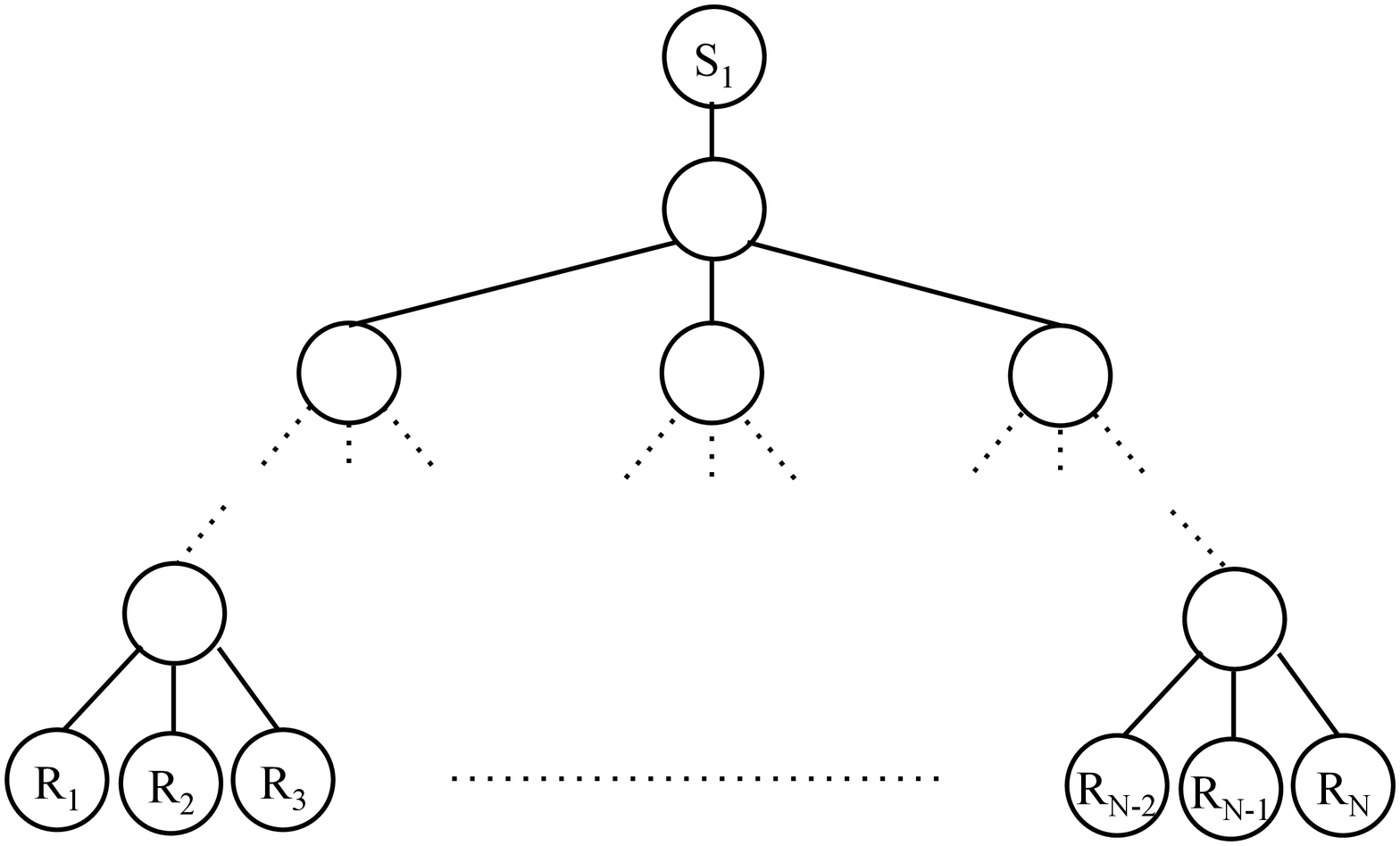}} 
\caption{\label{fig-examples} Four synthetic $G_{S_1\times\mathcal{R}}$ topologies used to evaluate the performance of Alg.~\ref{alg-gbs} (the GBS approach) in simulations.}
\end{figure*}

We denote the benefit of each type for a quartet $(R_i,R_j)$ by \textit{type1\_B}, $\cdots$, \textit{type4\_B} in Alg.~\ref{alg-gbs}, and define it as follows. Each quartet type limits the number of candidate edges where $J_i$ and $J_j$ can be located on, in the way depicted in Fig.~\ref{fig-2by2}. The benefit of a type for $(R_i,R_j)$ is the ratio of the number of edges where $J_i$ and $J_j$ can potentially be located on after learning this type, divided by the current number of candidate edges for the locations of $J_i$ and $J_j$. The worst case (minimum) benefit of $(R_i,R_j)$ results from the type for which this ratio is maximized, and the maximum of these worst case benefits over all quartets is given by the quartet with minimum ratio.

In order to provide an analytical upper bound on the number of quartets required by Alg.~\ref{alg-gbs}, one can try to use the main result of \cite{gbs-nowak}, which indicates that Alg.~\ref{alg-gbs} requires $\log_2 |\mathcal{H}|$ quartets.\footnote{This is the best case, where the constants $c^*$ and $k$ in \cite{gbs-nowak} are both as small as possible. In practice, there is an additional constant factor for $\log_2 |\mathcal{H}|$.} However, we cannot compute $|\mathcal{H}|$ exactly in our problem; we can only provide a loose upper bound on that, which is $N!$.\footnote{The bound is obtained by starting from the $S_1$ tree and considering all possible placements of $J_i$ on $P_{1i}$, $\forall$ $i$. Fig.~\ref{fig-examples}(c) shows that there are $N\times N\times (N-1)\cdots \times 2\cong N!$ possible such placements. In practice, the routing assumptions in Section~\ref{sec-statement} impose some constraints on possible $J_i$ locations. Also, the type of each quartet may rule out some types for the other quartets. Therefore, the exact $|\mathcal{H}|$ depends on the topology and we cannot compute it.} Therefore, we obtain the upper bound of $\log N! \approx N\log N$ on the number of quartets required by Algorithm~\ref{alg-gbs}, which is loose, and much larger than the lower bound. In the next section, we evaluate the performance of Alg.~\ref{alg-gbs} via simulation to obtain a better estimate of the number of quartets it requires to query in order to infer different topologies.

\subsection{Performance Evaluation}
\label{sec-evaluation}

\subsubsection{Simulation Setup}

\begin{figure}[t!]
\centering
\subfigure[A realistic $2$-by-$16$ topology from a US University departmental LAN \cite{merging}.]{\includegraphics[scale=0.21]{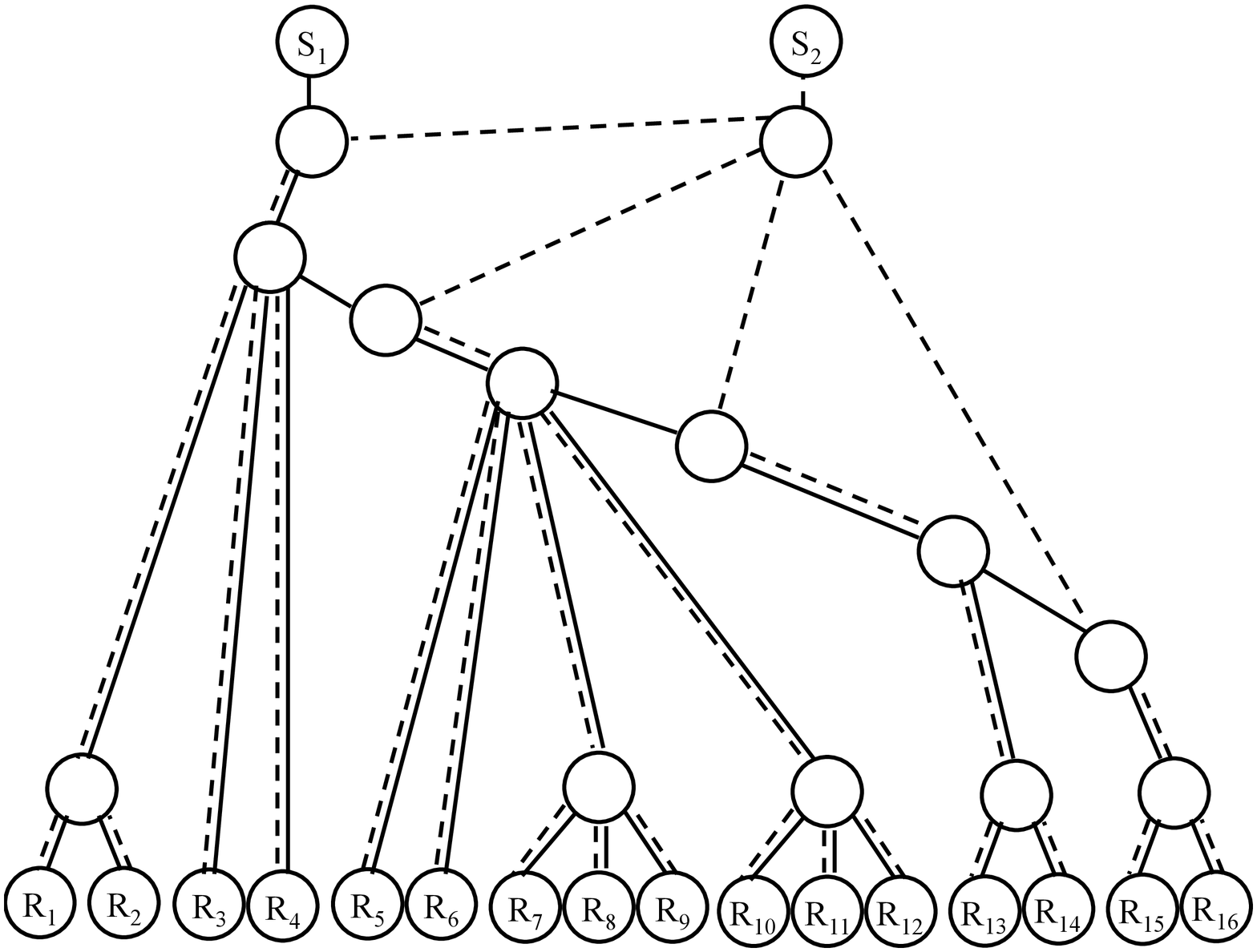}}
\subfigure[A $2$-by-$16$ topology generated from the Exodus topology \cite{rocketFuel}.]{\includegraphics[scale=0.21]{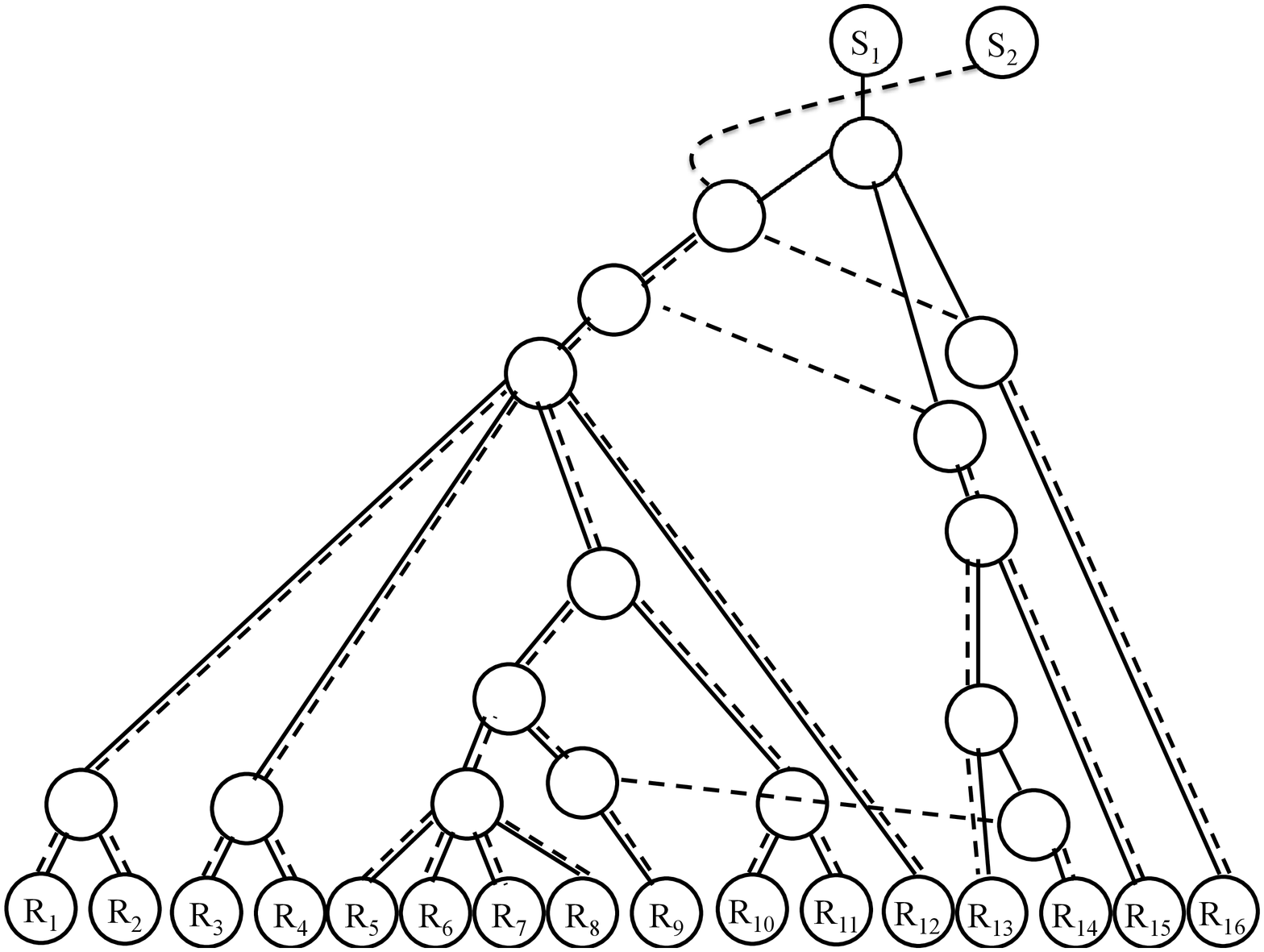}}
\caption{\label{fig-realistic} Two realistic $2$-by-$N$ topologies used to evaluate the performance of Alg.~\ref{alg-gbs} (GBS). The solid lines indicate the paths taken by probes from $S_1$ and the dashed lines indicate the paths taken by probes from $S_2$.} 
\end{figure}

We evaluate Alg.~\ref{alg-gbs} in simulations over both synthetic topologies (as shown in Fig.~\ref{fig-examples}) and realistic topologies (as shown in Fig.~\ref{fig-realistic}), and we compare it to the lower bound. The main performance metric of interest is the number of quartets queried in order to exactly infer the topology, which directly translates into measurement overhead. Additional metrics include the running time and the memory used by the algorithm, \ie the computational complexity.

For the synthetic topologies, we illustrate only the $1$-by-$N$ tree topology of $S_1$ in Fig.~\ref{fig-examples}. We consider the star topology, ``perfect'' and ``tall'' binary trees (referring to the topologies depicted in Fig.~\ref{fig-examples}(b) and \ref{fig-examples}(c), respectively), and perfect ternary trees, for the $G_{S_1\times\mathcal{R}}$ tree topology. Starting from this tree, we then create a $2$-by-$N$ topology, with sources $S_1$ and $S_2$, by choosing the location of each joining point $J_i$ (for receiver $R_i$) on a single logical link, selected uniformly at random, on $P_{1i}$ in $G_{S_1\times\mathcal{R}}$. For each $G_{S_1\times\mathcal{R}}$ in Fig.~\ref{fig-examples}, we consider 100 realizations of such random placements (resulting in different $2$-by-$N$ topologies)  and report the average number of quartets required for these topologies in the next section.

For the realistic topologies, we show the complete $2$-by-$N$ topology in Fig.~\ref{fig-realistic}. Fig.~\ref{fig-realistic}(a) depicts a US University departmental LAN with 16 receivers, first used in \cite{merging}. Fig.~\ref{fig-realistic}(b) is a $2$-by-$16$ directed acyclic graph (DAG), extracted from the Exodus topology, which is a large commercial ISP whose backbone map was inferred by the Rocketfuel project \cite{rocketFuel}. To generate this topology, we picked randomly two nodes of Exodus (nodes 5, 36) to be the sources, and selected all sixteen nodes to which both sources had routes to be the receivers. We then found the shortest path trees from each source to the receivers, and considered the overlap between these two trees. 

Our experiments are conducted using the Python implementation of Algorithm~\ref{alg-gbs}, which we have made available online \cite{code}. It takes as input any topology and returns the number of quartets required by Algorithm~\ref{alg-gbs} to infer that topology. Next, we summarize the simulation results. 

\subsubsection{Simulation Results (for the Number of Quartets)} 

When $G_{S_1\times\mathcal{R}}$ is a star topology as depicted in Fig.~\ref{fig-examples}(a),  Alg.~\ref{alg-gbs} always identifies the $2$-by-$N$ topology by querying only $\lceil \frac{N}{2} \rceil$ quartets, which is the lower bound. Therefore, it is optimal.

\begin{figure}[t!]
\centering
\includegraphics[width=8.0cm]{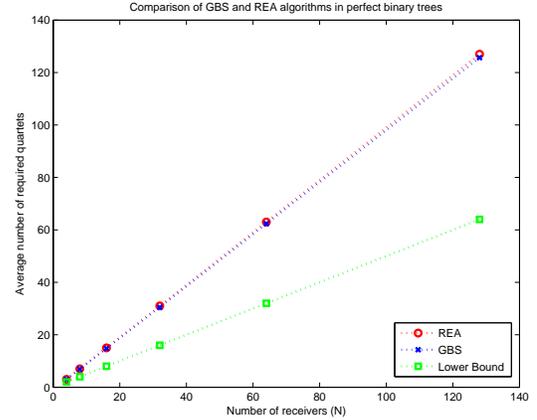}
\caption{\label{fig-binaryResults} Simulation results for the average number of quartets required by Alg.~\ref{alg-gbs} (GBS) to infer the $2$-by-$N$ topology when $G_{S_1\times\mathcal{R}}$ is a perfect binary tree (Fig.~\ref{fig-examples}(b)) of various sizes, $N=4,...,128$. The results are averaged over 100 realizations of random placements of the joining points. The standard deviation error bars (not shown) are comparable with the marker size. The figure also shows the number of quartets required by Alg.~\ref{alg-merging} (REA) and the lower bound in comparison to Alg.~\ref{alg-gbs} (GBS).}
\end{figure}

When $G_{S_1\times\mathcal{R}}$ is a perfect binary tree as shown in Fig.~\ref{fig-examples}(b), Alg.~\ref{alg-gbs} requires different numbers of quartets, between $\frac{N}{2}$ and $N$, in different $2$-by-$N$ topologies. However, as shown in Fig.~\ref{fig-binaryResults}, on average, Alg.~\ref{alg-gbs} requires $\sim N$ quartets. 

Similar results are obtained for tall binary trees (Fig.~\ref{fig-examples}(c)) and perfect ternary trees (Fig.~\ref{fig-examples}(d)). Here, we omit the figures and only report the results. When $G_{S_1\times\mathcal{R}}$ is a tall binary tree, the number of quartets required by Alg.~\ref{alg-gbs} varies depending on the quartet types in different $2$-by-$N$ topologies; however, in our simulations on tall binary trees with $N>100$ receivers, we observe that in at least $80\%$ of the realizations, Alg.~\ref{alg-gbs} requires $N-1$ quartets. This percentage increases up to $99\%$ in topologies with $N<100$ receivers. When $G_{S_1\times\mathcal{R}}$ is a perfect ternary tree, again on average, Alg.~\ref{alg-gbs} requires $N-1$ quartets, while for some topologies, it requires even more than $N$ quartets.

For the realistic topologies in Fig.~\ref{fig-realistic}(a) and Fig.~\ref{fig-realistic}(b), Alg.~\ref{alg-gbs} identifies both $2$-by-$16$ topologies by querying $14$ ($=N-2$) quartets.

Therefore, in our simulations, we find out that Alg.~\ref{alg-gbs} only performs as well as one could hope for, \ie it requires as few quartets as the lower bound, for flat $G_{S_1\times\mathcal{R}}$ topologies, such as the star topology in Fig.~\ref{fig-examples}(a). In other topologies, such as binary/ternary trees or realistic topologies, it requires many more queries, and each round of querying is extremely complex: at each step, Alg.~\ref{alg-gbs} needs to calculate the worst case benefits of all the quartets, in order to pick the best one among them. In fact, the time complexity of Alg.~\ref{alg-gbs} is $O(N^3)$, and its memory requirement is also high because it requires to keep track of all the benefits and the worst case benefits for all the quartets, as well as all the path updates for the location of each joining point, and so forth.

Since Alg.~\ref{alg-gbs} is not very efficient in practice as described above, we propose an alternate algorithm in the next section, which is much simpler and more efficient than the GBS approach.

\section{The Receiver Elimination Algorithm}
\label{sec-algorithm}

\begin{figure}[t!]
\centering
\includegraphics[scale=0.29]{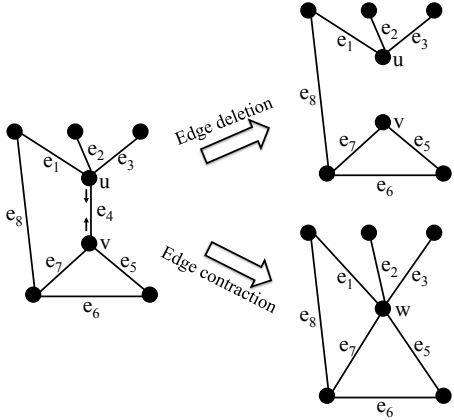}
\caption{Deletion and contraction of edge $e_4$ in a graph.} 
\label{fig-contraction}
\end{figure}

In this section, we design another greedy algorithm as an alternative to the GBS approach, called the Receiver Elimination Algorithm (REA), which requires more queries than GBS for some topologies, but each iteration is extremely simple and fast, and it scales linearly in the number of receivers. In particular, given $G_{S_1\times \mathcal{R}}$ and the ability to query the type of any quartet, REA is able to identify all $N$ joining points where $G_{S_2\times \mathcal{R}}$ merges with $G_{S_1\times \mathcal{R}}$, \ie the entire $2$-by-$N$ topology, in $N-1$ steps.

Let every edge $e$ in $G_{S_1\times \mathcal{R}}$ have a unique name: $label(e)$. In our algorithm, we use two operations ``edge deletion'' and ``edge contraction'', depicted in Fig.~\ref{fig-contraction} and defined as follows. 
\begin{definition}
{\em Deleting} edge $(u,v)$, entails taking that edge out of the graph while the end-nodes $u$ and $v$, and the labels of the remaining edges in the graph remain unchanged. 
\end{definition}
\begin{definition}
{\em Contracting} edge $(u,v)$ into node $w$, consists of deleting that edge and merging $u$ and $v$ into a single node $w$. The labels of the remaining edges do not change (although nodes may be renamed to $w$). 
\end{definition}

REA is described in Alg.~\ref{alg-merging}. It starts from the $S_1$ tree ($G_{S_1\times \mathcal{R}}$) and proceeds by selecting one quartet to query at each step (\ie two receivers $R_i, R_j$ to send probes to, from sources $S_1, S_2$). The two receivers ($R_i, R_j$) in the selected quartet are sibling leaves.  Based on the type of the selected quartet, Alg.~\ref{alg-merging} identifies exactly one joining point in one step. It then updates $G_{S_1\times \mathcal{R}}$ by deleting the receiver whose joining point has been identified and the last incoming edge to that receiver. That is why we call it the Receiver Elimination Algorithm. Furthermore, if a node of degree two appears in $G_{S_1\times \mathcal{R}}$ as a result of this edge deletion, the algorithm eliminates that node by contracting the corresponding edge. The algorithm continues iteratively until there is one edge left, \ie all joining points are identified. This way, Alg.~\ref{alg-merging} identifies all joining points (where paths from $S_2$ to each receiver join the $S_1$ tree), one-by-one, proceeding from the bottom to the root of the tree. Next, we describe an illustrative example.

\begin{algorithm}[t!]
\begin{footnotesize}
\caption{\label{alg-merging} REA starts from $G_{S_1\times\mathcal{R}}$, selects the quartets sequentially, queries their types, and merges them until identifying all joining points $\mathcal{J}_N$.}
\begin{algorithmic}[1]
\State Let $J$ be a vector of length $N$ of edge labels, which represents the locations of the joining points.
\While{$|\mathcal{R}|>1$}
\State Pick any two receivers $R_i$, $R_j$ in $G_{S_1\times \mathcal{R}}$, such that $R_i$ and $R_j$ are siblings; denote their parent by $P$.
\State Query the type of $(R_i,R_j)$.
\Switch{$(R_i,R_j)$}
\Case{type 1}
\State $J_i\equiv J_j$ 
\State Delete $R_i$ and edge $(P,R_i)$. 
\If{outdeg(P)==1}
\State Contract $(P,R_j)$ into $R_j$. 
\EndIf
\EndCase
\Case{type 2}
\State $J_j=label((P,R_j))$ 
\State Delete $R_j$ and edge $(P,R_j)$. 
\If{outdeg(P)==1}
\State Contract $(P,R_i)$ into $R_i$. 
\EndIf
\EndCase
\Case{type 3}
\State$J_i=label((P,R_i))$ 
\State Delete $R_i$ and edge $(P,R_i)$. 
\If{outdeg(P)==1}
\State Contract $(P,R_j)$ into $R_j$. 
\EndIf
\EndCase
\Case{type 4}
\State $J_j=label((P,R_j))$ 
\State Delete $R_j$ and edge $(P,R_j)$. 
\If{outdeg(P)==1}
\State Contract $(parent(P),P)$ into $P$. 
\EndIf
\EndCase
\EndSwitch
\EndWhile
\State /*There is one remaining receiver, which we call $R_{z}$.*/
\State Let $J_{z}=label((parent(R_{z}),R_{z}))$. 
\State Output $J$.
\end{algorithmic}
\end{footnotesize}
\end{algorithm}

\begin{figure*}[t!]
\centering  
\subfigure[The $G_{\mathcal{S}\times\mathcal{R}}$ topology, which we want to identify.]{\includegraphics[scale=0.2]{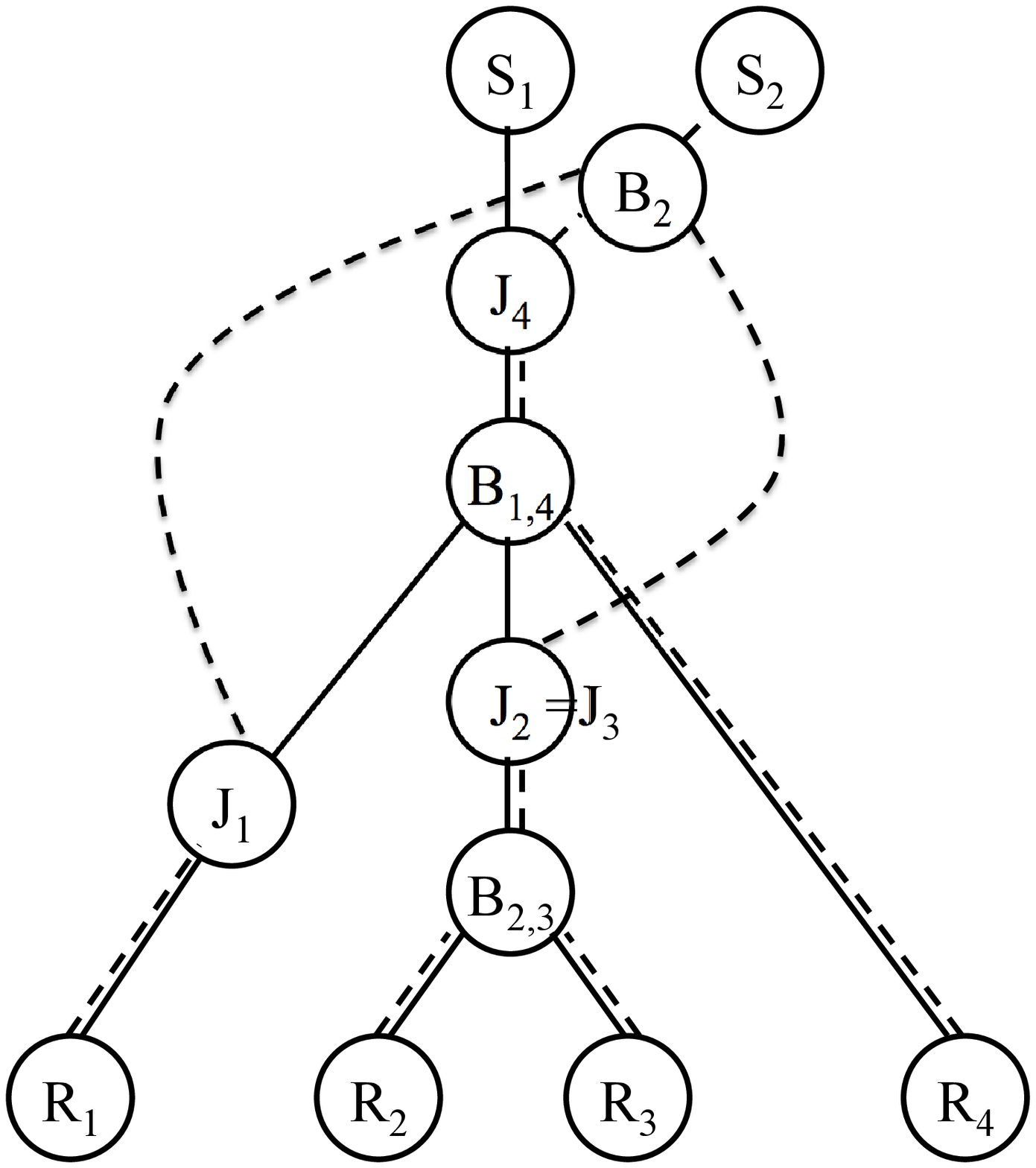}} 
\hspace{0.04cm}
\subfigure[$G_{S_1\times \mathcal{R}}$ ($T_4$). $(R_2,R_3)$ is of type 1; thus $J_2\equiv J_3$.]{\includegraphics[scale=0.2]{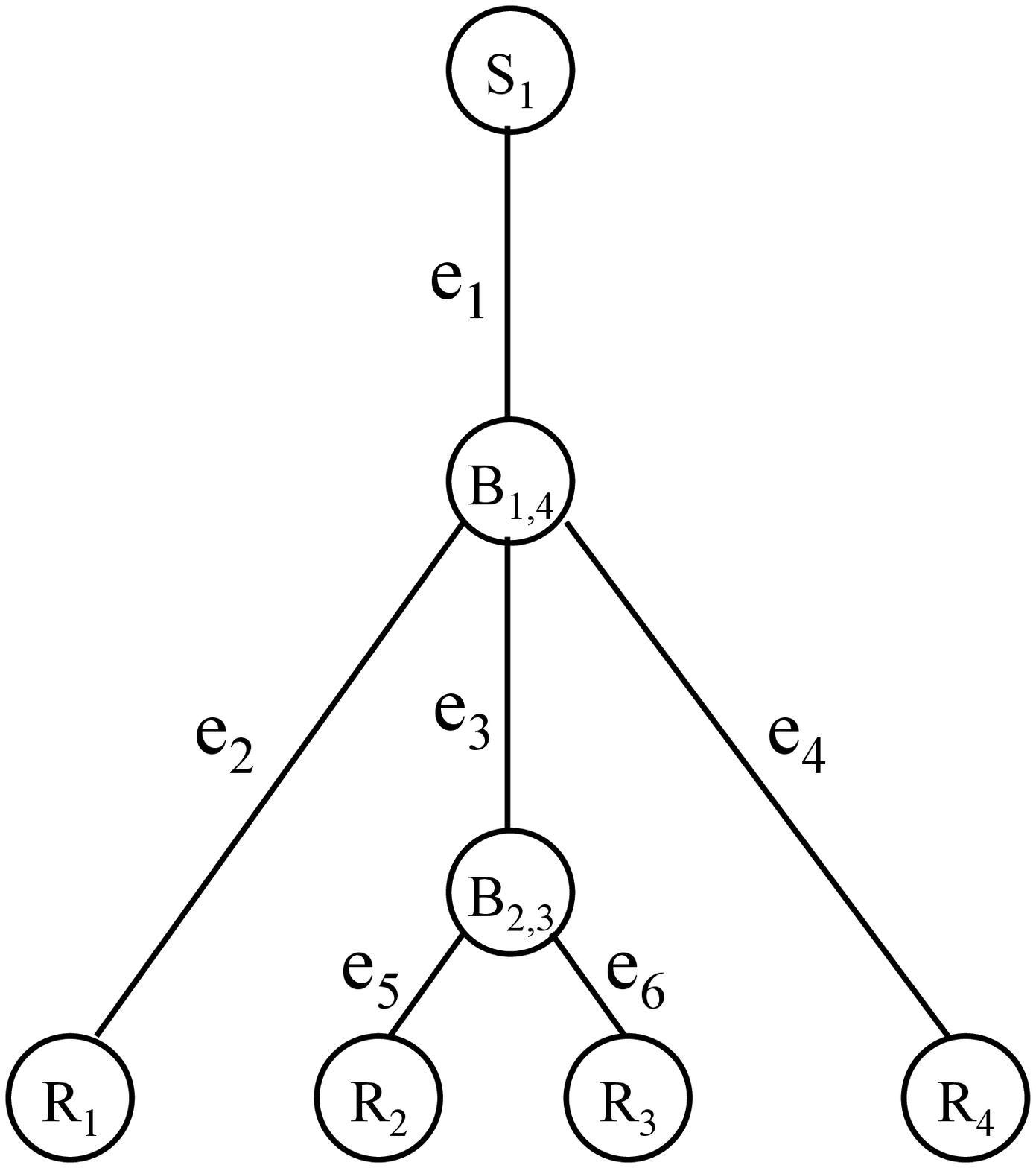}} 
\hspace{0.04cm} 
\subfigure[$T_3$. $(R_1,R_3)$ is of type 4; thus $J_3$ is identified on $e_3$.]{\includegraphics[scale=0.2]{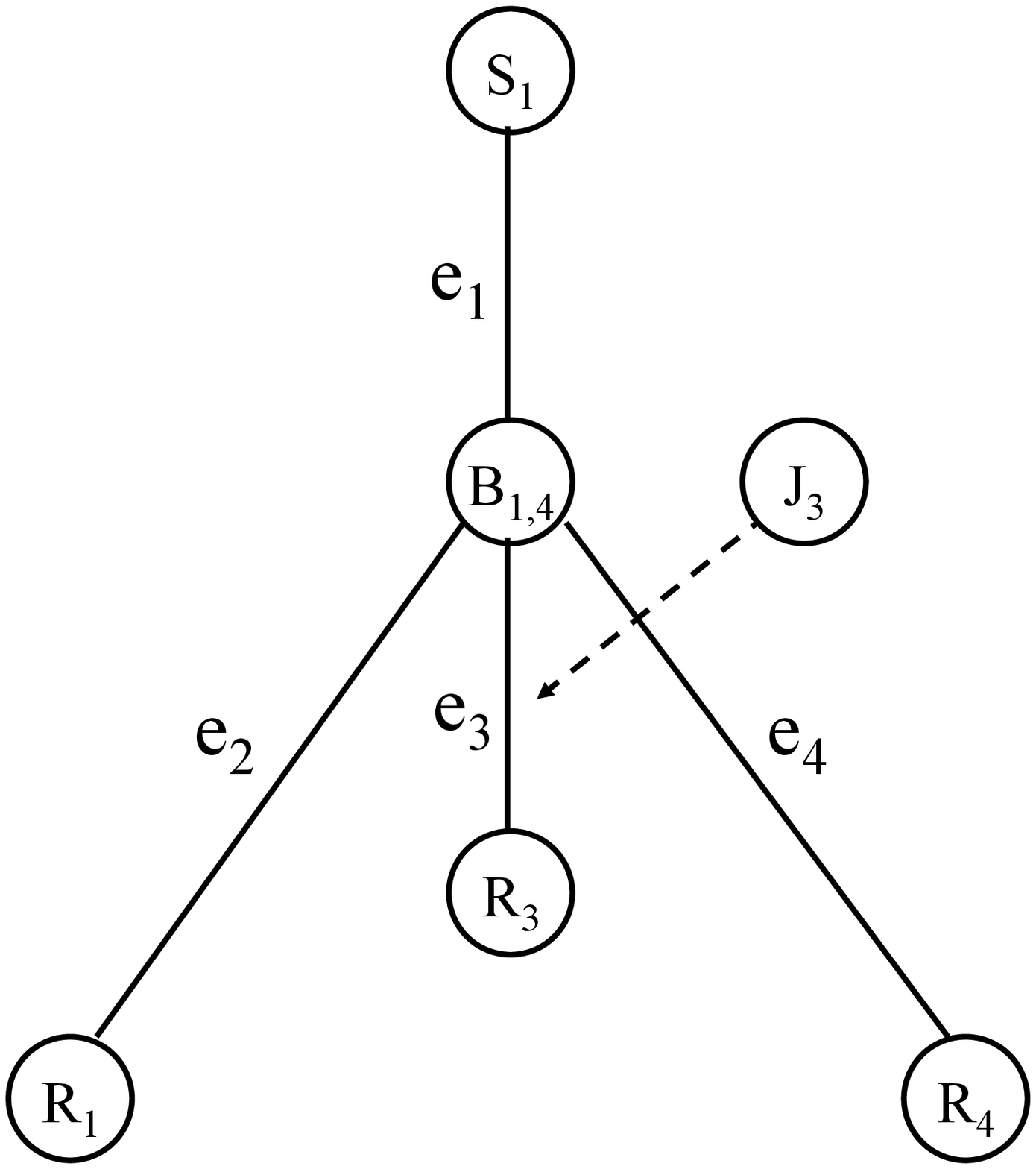}} 
\hspace{0.04cm} 
\subfigure[$T_2$. $(R_1,R_4)$ is of type 3; thus $J_1$ is identified on $e_2$.]{\includegraphics[scale=0.2]{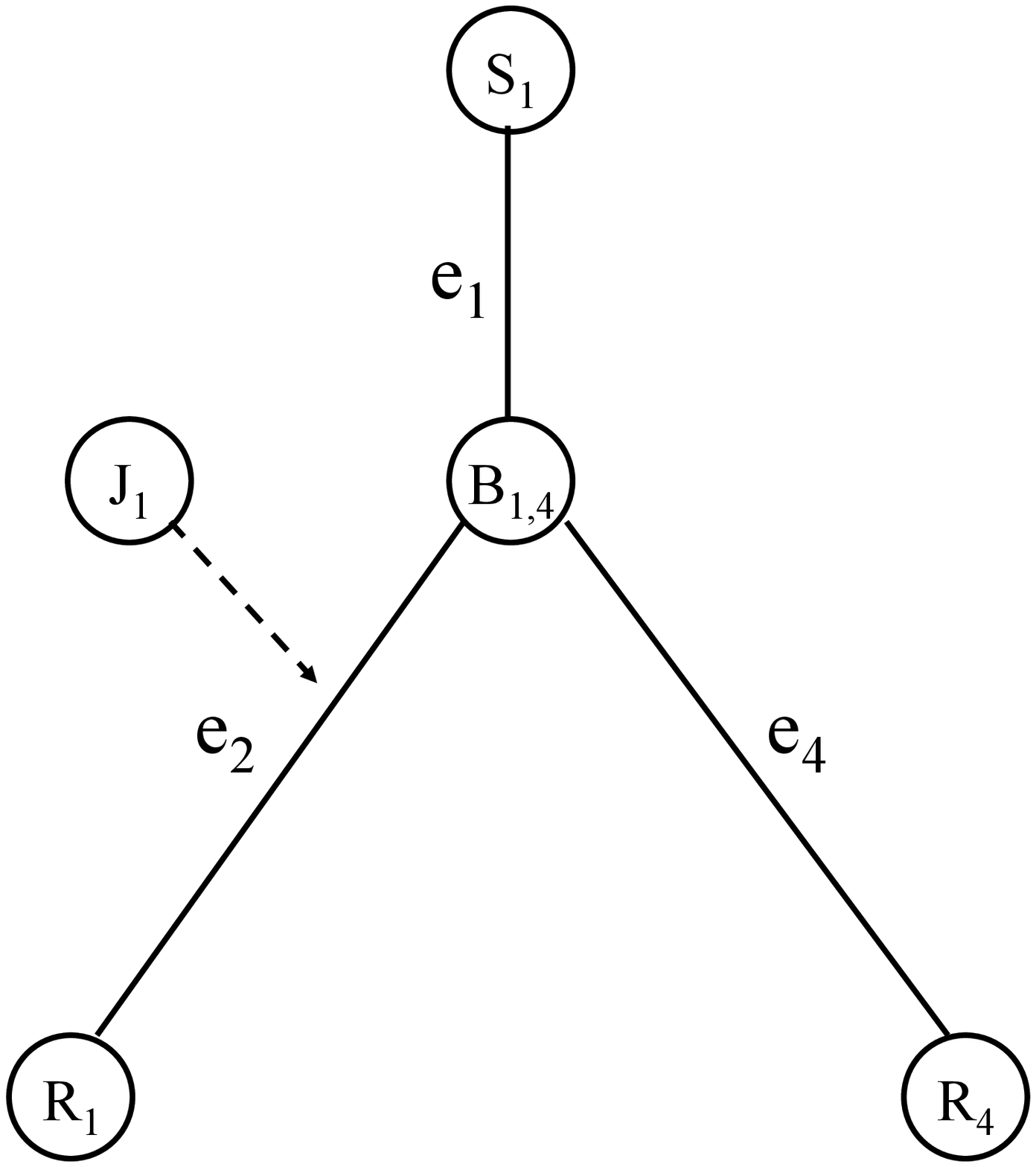}} 
\hspace{0.04cm} 
\subfigure[$T_1$. $R_{z}=R_4$; thus $J_4$ is identified on $e_1$.]{\includegraphics[scale=0.2]{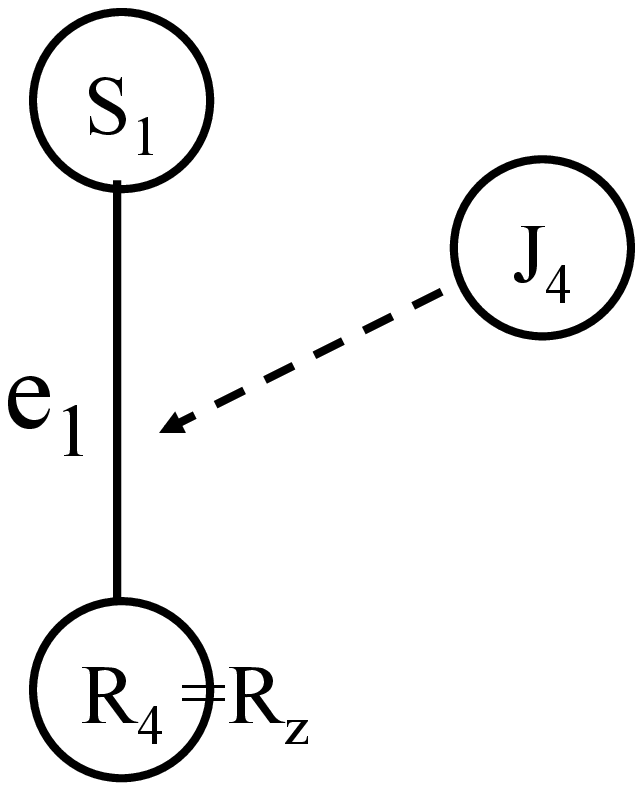}} 
\caption{The steps (b), (c), (d), and (e), performed by Alg.~\ref{alg-merging} (REA) to identify the $2$-by-$N$ topology in (a). The output of the algorithm is $J=[e_2,e_3,e_3,e_1]$.}
\label{fig-algExecution}
\end{figure*}

\begin{example}
Fig.~\ref{fig-algExecution}(b)-(e) demonstrate the steps performed by REA to identify the $2$-by-$N$ topology shown in Fig.~\ref{fig-algExecution}(a). The algorithm starts from $G_{S_1\times\mathcal{R}}$ shown in Fig.~\ref{fig-algExecution}(b); $e_1,...,e_6$ are the edge labels on this tree. The algorithm first selects $(R_2,R_3)$ and queries its type. Since the answer is type 1, the algorithm assigns $J_2\equiv J_3$, and deletes $R_2$ and $e_5$. Since the degree of $B_{2,3}$ becomes 2, the algorithm contracts $e_6$ into $R_3$. 

In the second step shown in Fig.~\ref{fig-algExecution}(c), REA selects two sibling leaves $(R_1,R_3)$, randomly out of three possible pairs of siblings, and queries its type. Since it is type 4, the algorithm identifies $J_3$ on $e_3$ (which, together with the previous step, means that $J_2$ is also identified). It also deletes $R_3$ and $e_3$. There is no contraction in this step as $B_{1,4}$'s degree is $>2$. 

In the third step shown in Fig.~\ref{fig-algExecution}(d), $(R_1,R_4)$ is selected and queried; it is of type 3. Therefore, the algorithm identifies $J_1$ on $e_2$,  deletes $R_1$ and $e_2$, and contracts $e_4$ into $R_4$. Since there is only one receiver left, there are no more quartets to query; thus the algorithm exits the while loop and proceeds to the last step (line 26). For $R_{z}=R_4$, the algorithm identifies $J_4$ on $e_1$, as shown in Fig.~\ref{fig-algExecution}(e). The identified joining points agree with the real locations in $G_{\mathcal{S}\times\mathcal{R}}$ topology in Fig.~\ref{fig-algExecution}(a), which demonstrates the correctness of the algorithm. 
\hfill{$\blacksquare$}
\end{example}

\subsection{Properties of REA}

Let $T_N = G_{S_1\times\mathcal{R}}$ denote the logical tree from $S_1$ to all $N$ receivers, which we assume to be known. In this section, we use the notation $T_N$ to emphasize that this initial tree $G_{S_1\times\mathcal{R}}$ contains $N$ receivers. After each iteration through the while loop in Alg.~\ref{alg-merging}, one receiver is deleted. We write $T_k$ to denote the tree (rooted at $S_1$) obtained at the end of iteration $(N-k)$, at which point there are $k$ receivers remaining. Let $\mathcal{J}_k$ denote the set of joining points, which still remain to be identified after iteration $(N-k)$, \ie one for each remaining receiver.

\begin{proposition}
Let $T_k$ and $\mathcal{J}_k$ be given. The next iteration of Alg.~\ref{alg-merging} (lines $3-25$) produces $T_{k-1}$ and $\mathcal{J}_{k-1}$, which satisfy the following properties:

1) The $S_1$ topology is still a logical tree, and it has $k-1$ receivers (\ie one receiver and its corresponding edge are deleted from $T_k$). Therefore, we denote it by $T_{k-1}$.

2) One joining point has been identified; therefore, the algorithm has $k-1$ more joining points in $\mathcal{J}_{k-1}$ to identify.

3) All joining points in $\mathcal{J}_{k-1}$ are located on edges in $T_{k-1}$.
\end{proposition}

\begin{IEEEproof}
These properties follow directly from the operations performed by one step of Alg.~\ref{alg-merging}:

1) In each iteration, a single receiver is eliminated from the tree. Consequently, the only node that can possibly have degree two (or out-degree one) after deleting the receiver is its parent, $P$. However, after each deletion, Alg.~\ref{alg-merging} tests to see if $P$ has out-degree 1, and if it does, then an additional contraction is performed so that the resulting tree, $T_{k-1}$, is still logical. 

2) When $(R_i,R_j)$ is of type 2, 3, or 4, we can see in lines 12, 17, and 22 of the algorithm, respectively, that one joining point is identified. When $(R_i,R_j)$ is of type 1, line 7 assigns to $R_i$, the same joining point as $R_j$'s. Then, in line 8, $R_i$ is  deleted so that we do not create a loop by assigning $J_i$ again to $J_j$ later. Also, $J_j$ eventually becomes identified, either in one of the other types (2, 3, or 4) in the while loop, or in the last line of the algorithm. Thus, we have $\mathcal{J}_{k-1}$ after one step.

3) Alg.~\ref{alg-merging} changes $T_k$ by 2 processes: edge deletion and edge contraction. We show that neither deletion nor contraction can eliminate an edge in $T_k$ that contains a joining point in $\mathcal{J}_{k-1}$. 

{\em Deletion:} Alg.~\ref{alg-merging} is constructed such that any edge deleted from the $S_1$ tree contains either no joining point (if $(R_i,R_j)$ is of type 1) or exactly one joining point, corresponding to the receiver being removed along with that edge (if $(R_i,R_j)$ is of type 2, 3, or 4).

{\em Contraction:} An edge is contracted only when it does not contain any joining point, neither for $R_i$ and $R_j$ (see lines $9-10$ for type 1, lines $14-15$ for type 2, lines $19-20$ for type 3, and lines $24-25$ for type 4), nor for any other receivers (since $(R_i,R_j)$ are sibling leaves, the contracted edge cannot contain any joining point for any other receiver.\footnote{Algorithm~\ref{alg-merging} selects {\em sibling} receivers $R_i$ and $R_j$ at each step. Therefore, there are only two potential candidates for the joining points that can be identified at this step: $J_i$ and $J_j$.}). 
\end{IEEEproof}

The following theorem establishes the correctness and complexity of Algorithm~\ref{alg-merging} (REA).
\begin{theorem}
\label{thm-Nsteps}
REA terminates in $N$ steps and correctly identifies all $N$ joining points after querying $N-1$ quartets.
\end{theorem}
\begin{IEEEproof}
The proof is via induction. In the beginning, $T_N=G_{S_1\times\mathcal{R}}$ is a logical tree and according to Corollary 1 in \cite{journal}, the joining points are identifiable using sufficient quartets. Our inductive step is one iteration of the while loop. First, note that there exist two sibling receivers at every step: it is enough to pick one of the lowest receivers (\ie a receiver with the largest distance from the source); it will always have a sibling because of the logical tree topology.  The algorithm queries one quartet per step, identifies one joining point per step, and at the end of the step, it preserves properties 1, 2, and 3. The while loop terminates in $N-1$ iterations and there is one additional step for $R_{z}$ after the loop (which does not use any quartet). Therefore, the algorithm terminates in $N$ steps, and correctly identifies all $N$ joining points by querying exactly $N-1$ quartets. 
\end{IEEEproof}

{\em Discussion.} An important observation is that the $N-1$ quartets are not known a priori, but are easily selected in a sequential way, as needed; this makes REA easy to implement in practice using active probing. Another observation is about the running time: exactly $N-1$ quartets need to be queried (by sending sets of probes). This is much less than the $\binom{N}{2}$ possible quartets queried by a brute-force approach \cite{merging, journal}, but higher than the lower bound on the number of required quartets by any algorithm ($\lceil \frac{N}{2} \rceil$, Theorem~\ref{theorem-minimum2by2s}). Therefore, REA is not optimal, but it is simple, efficient, and provably correct. The next section compares the performance of REA to GBS in different topologies.

\subsection{Comparison to GBS}
\label{sec-comparison}

In Section~\ref{sec-evaluation}, we evaluated the performance of Alg.~\ref{alg-gbs} (GBS) in simulations over both synthetic topologies of Fig.~\ref{fig-examples} and realistic topologies of Fig.~\ref{fig-realistic}. In this section, we compare the performance of Alg.~\ref{alg-merging} (REA) against Alg.~\ref{alg-gbs} (GBS) and the lower bound, over the same topologies. The performance metrics of interest include the number of quartets queried in order to exactly infer the topology, \ie the measurement overhead, as well as the running time and the memory used by each algorithm.

\subsubsection{The Number of Quartets} 

When $G_{S_1\times\mathcal{R}}$ is a star topology as in Fig.~\ref{fig-examples}(a), we saw in Section~\ref{sec-evaluation} that GBS is optimal and requires only $\lceil \frac{N}{2} \rceil$ quartets. Therefore, it performs better than REA, which requires $N-1$ quartets. 

On the other hand, when $G_{S_1\times\mathcal{R}}$ is a perfect binary tree as in Fig.~\ref{fig-examples}(b), we can see in Fig.~\ref{fig-binaryResults} that on average, REA performs very close to GBS, while GBS is much more complex than REA. Similar results are obtained for tall binary trees (Fig.~\ref{fig-examples}(c)) and perfect ternary trees (Fig.~\ref{fig-examples}(d)). As described in Section~\ref{sec-evaluation}, for both $G_{S_1\times\mathcal{R}}$ topologies, on average, REA performs close to GBS, and for some topologies, GBS requires even more than $N$ quartets. 

For the realistic $2$-by-$16$ topologies in Fig.~\ref{fig-realistic}(a) and Fig.~\ref{fig-realistic}(b), we saw in Section~\ref{sec-evaluation} that GBS requires $N-2=14$ quartets, while REA requires $N-1=15$ quartets. 

Therefore, one can see that GBS only requires significantly fewer quartets than REA for flat $G_{S_1\times\mathcal{R}}$ topologies, such as the star topology in Fig.~\ref{fig-examples}(a). In other topologies, such as binary/ternary trees or realistic topologies, REA is preferred over GBS, because it is much simpler and it identifies the joining points using the same number of quartets (or even fewer quartets in large topologies) as GBS (\ie $N-1$). 

\subsubsection{Time and Space  Complexity}
\label{sec-complexity}

The time complexity of REA ($O(N)$) is significantly lower than that of GBS ($O(N^3)$). The reason is that at each step, REA only needs to select a pair of sibling receivers (any of them will do); while GBS calculates the worst case benefits of all the quartets, in order to pick the best one among them, which takes much longer. As an example, for a single realization of our simulations when $G_{S_1\times\mathcal{R}}$ is a perfect binary tree with $128$ receivers, the running time of REA is only $<1$ second, while that of GBS is $19$ seconds. This is a big difference when we consider a large number of realizations as described in Section~\ref{sec-evaluation}. 

The memory requirement of REA is also much lower than that of GBS. The reason is that REA only requires to store the (modified version of the) graph at each step; while GBS requires to keep track of all the benefits and the worst case benefits for all the quartets, all the path updates for the location of each joining point, and so forth.

\section{Extensions}
\label{sec-discussion}

In this section, we briefly outline the possible extensions to the active learning algorithms we have discussed so far.

\subsection{Extension to $M$-by-$N$ Topologies}

So far, we have focused on inferring a $2$-by-$N$ topology, which is a special but important case. $M$-by-$N$ topologies can be inferred by merging the tree topologies of the remaining $M-2$ sources to this $2$-by-$N$ topology, one source at a time. Assume that we have inferred a $k$-by-$N$ topology, $2\leq k< M$. To add the $(k+1)^{th}$ source, we need to identify each joining point of the new source, $S_{k+1}$, and any one of the $k$ sources in the current topology, $S_{i}$, $1\leq i\leq k$, for each receiver, on a single logical link in the $k$-by-$N$ topology (defined by all the branching points). Therefore, we need to apply REA (or GBS) to $S_{k+1}$ and any one (in the best case) or all (in the worst case) of the current $k$ sources. Therefore, for example using REA, the number of quartets required to identify the $M$-by-$N$ topology is between $(M-1)(N-1)$ and $\binom{M}{2}(N-1)$.

\subsection{Extension to Noisy Case}

So far, we have considered the noiseless scenario, where the answer to each query is the correct quartet type. One can extend the algorithms to deal with noisy queries, using the two approaches proposed in \cite{gbs-nowak}. The first one is a simple solution that applies to both GBS and REA; it repeats the query multiple times and considers the majority vote as the answer to that query. The second approach is more sophisticated and fits naturally in the GBS framework.\footnote{A similar solution for REA would be to perform the deletions and contractions probabilistically.} It assigns weights to each hypothesis using a probability distribution. The initial weighting is uniform, but it gets updated after each query. The update naturally boosts the probability measure of the hypotheses that agree with the answer to the query. At the end, the hypothesis with the largest weight is selected. We can adopt this approach for the GBS algorithm by incorporating the probability measures in the path updates and in computing the benefits. Using this approach, GBS can handle the noisy queries more naturally than REA. The query complexity and the probability of error for both approaches have been analyzed in \cite{gbs-nowak}.

\section{Conclusion}
\label{sec-conclusion}

Although active topology inference is a well-studied problem, to the best of our knowledge, this paper is the first to focus on efficient merging algorithms. We formulate the problem as multiple hypothesis testing and develop an active learning algorithm based on GBS. We also propose an efficient Receiver Elimination Algorithm that queries only $N-1$ quartets, which is much less than the $\binom{N}{2}$ possible quartets. Furthermore, comparing it to the GBS algorithm in simulations, we find out that the simple REA is near-optimal, and comparable to the GBS approach in terms of the number of queries (thus measurement bandwidth), while having much lower time and space complexity. Therefore, it is preferable for all practical purposes.



\end{document}